\documentclass[conf]{new-aiaa}
\usepackage[utf8]{inputenc}
\usepackage{graphicx}
\usepackage{amsmath}
\usepackage[version=4]{mhchem}
\usepackage{siunitx}
\usepackage{longtable,tabularx}
\usepackage{subcaption}
\usepackage{mwe}
\usepackage[section]{placeins}
\usepackage{comment}
\usepackage{multirow}

\title{Applications and extensions of conditional space-time POD}
\title{Conditional space-time POD extensions for stability and prediction analysis}

\author{Spencer L. Stahl \footnote{Graduate Research Assistant, Mechanical and Aerospace Engineering Department, The Ohio State University, AIAA Student Member.}}
\author{Chitrarth Prasad \footnote{Postdoctoral Researcher, Mechanical and Aerospace Engineering Department, The Ohio State University, AIAA Member.}}
\author{Hemanth Goparaju \footnote{Graduate Research Assistant, Mechanical and Aerospace Engineering Department, The Ohio State University, AIAA Student Member.}}
\author{Datta Gaitonde \footnote{Professor, Mechanical and Aerospace Engineering Department, The Ohio State University, and AIAA Fellow.}}
\affil{Mechanical and Aerospace Engineering, The Ohio State University, Columbus, OH 43210, USA}
\usepackage{setspace}

\begin{document}
 
\maketitle

\begin{abstract}
The correlation and extraction of coherent structures from a turbulent flow is a principle objective of data-driven modal decomposition techniques. 
The Conditional space-time Proper Orthogonal Decomposition (CPOD) offers insight into transient dynamics, revealing the causation of specific flow phenomenon - or events, in a customizable manner.
This work exploits the temporal evolution of CPOD modes in a reduced subspace, resulting in new extensions and adaptations that meet or exceed the capabilities of other decomposition methods.
Chiefly, it is demonstrated that the subsequent application of dynamic mode decomposition (DMD) to CPOD modes, provides a flexible tool to investigate targeted flow instabilities, both tonal and convective in nature.
By extending the CPOD time-horizon to educe the former type, it is shown that CPOD-DMD can exactly reproduce Spectral POD modes.
Regarding the latter, a multi-resolution framework (CPOD-mrDMD) yields a refined "cause and effect" stability analysis, capable of diagnosing the natural forcing mechanisms within the flow, and the resulting unstable modes.
In a separate application, CPOD properties are appreciated in the context of reduced order models, with an example of real-time flow prediction of extreme events  derived from an active sensor correlated to a CPOD mode.
The various CPOD functions and perspectives in this work are demonstrated on: the nonlinear chaotic Lorenz system, 3D intermittent turbulent spots in supersonic boundary layer transition, Schlieren video processing of unstarted inlet buzz, the aeroacoustic feedback forcing of a resonating impinging jet, and prediction of intermittent bluff-body wake structures impinging on a channel wall.

\end{abstract}

\section{Introduction}\label{sec:Intro}
Turbulent flows evolve with a range of spatial and temporal scales that result in complex nonlinear behavior and stochastic events. 
The fluid dynamics community has adopted multiple data-driven modal decomposition techniques that extract coherent structures within the turbulence to give insight into the underlying dynamics \cite{Taira2017a_modal}. 
Proper Orthogonal Decomposition (POD) is the powerful reduction tool at the center of these techniques, separating spatial correlations and temporal coefficients across a set of modes \cite{Berkooz1993}.
However, often lost in the \textit{correlation} of the resulting modes is the \textit{causation} of flow dynamics.
The generalized space-time POD, introduced by Lumley \cite{Lumley1967structure,Lumley_POD} before the ubiquity of computational analysis, addresses this issue with modes that evolve coherent structures over time. 
This orthogonality between space and time has been exploited in other decompositions, most commonly in the \textit{frequency} domain of Spectral POD (SPOD) \cite{Gordeyev2000,Citriniti2000} or the concomitant Dynamic Mode Decomposition (DMD) \cite{Schmid2010}, the latter of which is only temporally orthogonal.
Recently, Conditional space-time POD (CPOD \footnote{Conditional space-time POD has been abbreviated to CST-POD elsewhere \cite{hack2021extreme}, but is shorten to CPOD here for simplicity when combining with other methods (ex. CPOD-mrDMD). }) has been reintroduced as a tool to explore the \textit{transient} dynamics caused by a specified event in the flow \cite{Schmidt2019,hack2021extreme}.
This paper seeks to expand the functionality of CPOD to educe a variety of flow phenomena with connections to POD, DMD, and SPOD.
Furthermore, new CPOD-based analyses are proposed, such as a "cause and effect" stability framework and a CPOD based strategy to predict future flow events.

\begin{figure}
\centering
\includegraphics[width=.8\textwidth]{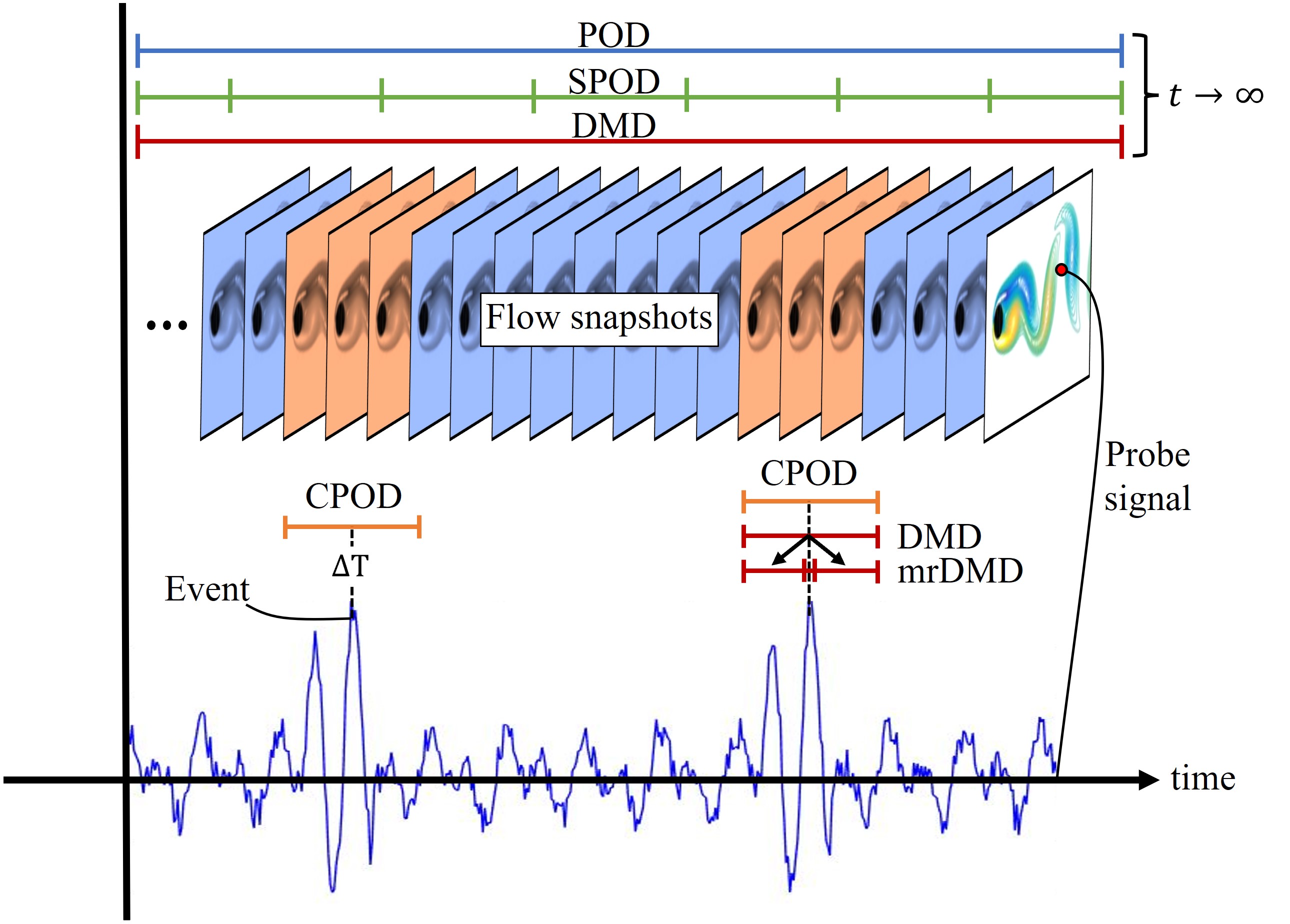}
\caption{Application of different data-driven modal decompositions. CPOD is applied to snapshot sequences of local event realizations while POD, SPOD, and DMD are typically applied over an entire statistically stationary data-set.  }
\label{fig:modal_overview}
\end{figure}

Before examining what space-time CPOD is in detail, it is necessary to first understand the role it plays apart from other popular fluid decomposition methods.
An overview comparing the applicability of these decompositions is presented in Fig.~\ref{fig:modal_overview}.
To start, we consider only data-driven decompositions applied to flow snapshots and distinguish between approaches that act over an entire statistically stationary data-set ($t \rightarrow \infty$), and those that capture transient growth over a shorter time horizon ($\Delta T << \infty$).
Regarding the former type, common methods include space-only POD (or snapshot POD), SPOD, and DMD, of which the last two provide frequency information.
SPOD inputs ensembled data from multiple overlapping blocks of flow snapshots and decomposes a cross-spectral density tensor, producing energy ranked modes that correspond to a single frequency \cite{Towne2018_relation,Schmidt2020} .
On the other hand, DMD approximates a best fit linear operator that optimally describes the dynamics across the entire flow sequence~\cite{Rowley2009,Schmid2010,Tu2014}.
When applied to shorter transient data-sets, DMD yields mode frequencies with now relevant growth rates, making it an effective tool for stability analysis.
Multi-resolution DMD (mrDMD)~\cite{Kutz2015} provides a temporally local framework and recursively performs DMD to elucidate multi-scale features; these properties will be co-opted in this work for revealing causality of flow instabilities.
Uniquely, CPOD is applied to a statistically stationary data-set, but extracts short-term sequences of several uncorrelated event realizations that are conditionally selected.
The decomposition of the ensembled event sequences produce a set of CPOD modes optimally ranked by energy (variance) that are coherent over space and a finite time window \cite{Schmidt2019}.
This effectively evolves the average event sequence and isolates flow structures in higher-order modes.

CPOD is closely relate to the simpler conditional average (or sample average), which is also well suited to characterize time-local events \cite{Antonia1981_cond_samp_rev,Kamine2016,Bogey2019_temporal_jet} and shown to have modeling advantages over the previous decomposition techniques \cite{Akamine2020}.
For example, the definition of the "event" to be extracted from conditional averaging is arbitrary, and can range from regularly occurring flow patterns to intermittent processes that induce rare instabilities \cite{Schmidt2019,Brunton2017_rare}.
All that is required for the CPOD algorithm is a conditional criteria to identify the instances in time when the process of interest is taking place.
Because of this, the choice of conditional criteria is highly customizable to isolate any phenomenon. 
Multiple strategies of conditional event selection are given throughout this work, usually in the form of a simple threshold function that catalogs abnormally large fluctuations in a probe signal, however more targeted methods that require precursor calculations have been used effectively \cite{stahl2021_aviation,hack2021extreme}.
CPOD can also be used diagnostically without \textit{a priori} knowledge of what the event is.
For example, the cause behind an extreme fluctuation (of say pressure), can be investigated with CPOD by identifying all large fluctuations at the location of the anomaly.
In this work, flow events include the intermittent turbulent spots of a supersonic boundary layer, supersonic unstarted inlet buzz, acoustic feedback resonance instabilities, and the wake behind a bluff body that interacts with a channel wall.  

Flow instability, such as that encountered in the above examples or induced by external forcing, is a major area of research related to decomposition methods \cite{Schmid2007,Taira2017a_modal,Theofilis2011}.
In general, the objective of stability analysis is to characterize the evolution and growth of small perturbations applied to a base-flow or equilibrium state.
Therefore, stability methods fundamentally require input of a solution to the Navier-Stokes equations - typically with a prescribed forcing, and output eigenvectors corresponding to unstable modes of which energy growth (gain) can be quantified.
Multiple equation-based and data-driven strategies exist for this task \cite{Theofilis2003,Luchini2014_adjoint,Ranjan2020_mfp,Herrmann2021}, with most being a form of Linearized Stability Theory (LST) \cite{Schmid2000_LST,Gomez2012}.
While traditional methods have proven successful in parameterizing flow sensitivities and mapping instability bifurcations \cite{Schmid2014}, fully educing the nonlinear turbulent nature of instabilities still remains a challenge, particularly for high Reynolds number flows \cite{Taira2020_modal_b}. 
This is partially remedied by data-driven stability approaches, which usually apply DMD to the perturbation evolution to solve the eigenvalue problem \cite{Schmid2010,Goparaju2021,Ranjan2020_mfp,LeClainche2018}; DMD will be used in a similar capacity with the CPOD stability analysis introduced next.

CPOD provides an additional tool to investigate flow instabilities with a unique perspective. 
In particular, the appeal of CPOD lies in its ability to temporally and spatially characterize specific large fluctuation events, which often include the forcing mechanism and resulting instability within the CPOD mode itself; such forcing events are only approximated by stochastic or prescribed perturbations in other methods \cite{Ranjan2020_mfp,Schmid2000_LST,Towne2018_relation}.
Whereas LST isolates the linearized perturbation response around a laminar or mean base-flow, CPOD more realistically captures the event instability in context of the "natural" flow conditions that create the forcing event.
This is because the nonlinear dynamics throughout the global system are inherently accounted for in the transient CPOD modes.
For example, Schmidt \textit{et al.} \cite{Schmidt2019} investigated the instability associated with intermittent Mach waves radiating from a turbulent jet.
The forcing "seed" (initial snapshot) of the calculated CPOD mode was compared to the optimal initial condition of a linearized transient growth Navier-Stokes calculation that produced the same acoustic burst.
The excellent agreement between the evolution of that result and the CPOD mode, support continued use of CPOD for stability experiments.  
In section~\ref{secn:CPOD_DMD}, a purely data-driven stability analysis is posed using a subsequent application of DMD to a CPOD mode (CPOD-DMD).
Furthermore, a CPOD-mrDMD variant provides a refined "cause and effect" stability analysis by segregating the CPOD mode timeline before and after the event, thus isolating the dynamics of the forcing and response \cite{stahl_caes2022}, comparable to the goal of resolvent analysis \cite{Jeun2016_io}.

In this work, a resonating supersonic impinging jet is used to instantiate the stability extension of CPOD-DMD.
The impinging jet provides a great example of how CPOD-DMD can elucidate both absolute (resonance tones) and non-modal convective instabilities generated from a self-sustained acoustic feedback forcing.
Ideally, the resonating jet is also a well-studied system \cite{Jiang2019,Edgington-Mitchell2019_review} that has been explored with a variety of decomposition techniques, some of which are now referenced. 
Weightmen \textit{et al.} used POD to study the phase lag component of the feedback loop \cite{Weightman2017} and explore the effects of the nozzle shape on the acoustic receptivity process \cite{weightman_2019}. 
Karami \textit{et al.} studied the full non-linear unsteady evolution of round impinging jets from a Large Eddy Simulation (LES) \cite{Karami2020_characteristics}, applied SPOD to correlate structures to impinging tones \cite{Karami2018a_spod}, and also performed a linearized perturbation analysis to study the response of the acoustic feedback receptivity process  \cite{Karami2018,Karami2020_recept,Karami2021_nozz}.
Mendez~\textit{et al.} developed a multi-scale POD (mPOD) method which was tested on an impinging jet, yielding better convergence than Fourier or DMD modes \cite{Mendez2019}. 
Hildebrand \textit{et al.} performed a global stability analysis on another supersonic impinging jet, comparing the direct and adjoint modes to understand non-normal convective instabilities \cite{Hildebrand2017}.
Gojon and Bogey simulated, decomposed, and compared impinging tones with theoretical feedback models for round \cite{Gojon2016_azim,Gojon2017,Gojon2018,Bogey2017_feedback} and planar \cite{Gojon2016_planar,Gojon2019_planar} impinging jets, similar to the configuration studied here.

The functionality of CPOD is expounded upon in comparison to the decomposition methods in the aforementioned impinging jet references. 
Addressing those studies, the unprocessed computational solution or experimental data provide the most complete physics, but the acoustic feedback dynamics of interest are obscured by the turbulent nature of the jets.
POD greatly simplifies the flow with correlated structures that elucidate the impinging jet modeshapes, but the POD modes are not coherent over time. 
SPOD filters these modal structures at a particular frequency, however the monochromatic SPOD modes are void of transient dynamics contributed from other frequencies.
Conversely, the non-modal LST analyses isolates the transient dynamics, but does not capture all of the non-linear physics. 
CPOD has been applied to the impinging jet problem \cite{stahl_sci2022}, rectifying these shortcomings with limited drawbacks.
First, the non-linear evolution of the flow is inherently captured in CPOD modes, while also possessing the ability to separate and rank modes by energy, as with POD and SPOD.
Secondly, the targeted instability of a prescribed LST forcing can be analogously constructed using an intelligent CPOD event selection, and likewise decomposed with DMD as previously discussed.
Lastly, because CPOD modes are coherent over space and time, any periodicity within the time-horizon can be deduced.
Due to this property, it will be shown that CPOD-DMD can exactly recover global SPOD modes as well as local convective instabilities.. 
This relationship is explored in detail in Section~\ref{secn:CPOD_DMD}.

Finally, CPOD is discussed in the context of providing effective reduced order models (ROMs) for flow prediction and potentially control.
Most commonly, POD or DMD modes are the chosen ROM basis functions \cite{rowley2017_control} in combination with methods like Galerkin projection \cite{BerndRNoack} or Kalman filters \cite{Rowley2005}.
For example, atmospheric weather predictions dating back to the early work of Lorenz \cite{lorenz1956empirical} were modeled using Empirical Orthogonal Functions (a form of POD used in climate science), while recent work used POD from simulation data to predict real-time turbulence at an airport \cite{Kikuchi2015}.
Gordeyev \textit{et al.} developed a Temporal POD (TPOD) methodology, similar to CPOD, with the goal of characterizing the transient behavior between a nominal wake flow and a controlled flow state \cite{Gordeyev2013}.
Groups of trajectories from one initial state to another controlled state were categorized and effectively modeled with the TPOD results. 
The low-rank dynamics encapsulated in CPOD modes share similar properties and have potential applications in event prediction.
This consideration is investigated in Section~\ref{secn:predict} to predict the intermittent impingement of wake-flow structures behind a bluff body onto a nearby surface.
Here, a sensor on the bluff-body actively correlates its time-history with a CPOD mode that characterizes the wake impingement, yielding a near-term prediction certainty for future events. 
While this is a simplified prediction scheme to fit within the scope of the paper, it highlights the broader versatility of CPOD as a ROM and future avenues of research.

\section{Conditional Proper Orthogonal Decomposition}
\subsection{Theory}

The conditional space-time POD is formulated to optimally capture the variance of an ensemble of event realizations.
The mathematical theory presented follows the formulation of Schmidt \textit{et al.}\cite{Schmidt2019}.
The selection of the conditional criteria used to define the events and corresponding flow sequences $\textbf{q}_{r}(\textbf{x},t)$ is arbitrary; conditional methods will be discussed later. 
Here, $\textbf{q}$ is the flow-field sequence of event realization $r$, spanning the spatial and temporal domain $\textbf{x}$ and $t$ over the short-time window $\Delta T$.
Starting from the conditional expectation (sampled average) $E\{\cdot\}$ of each event and the space-time inner product $\langle \cdot,\cdot \rangle_{\textbf{x},t}$  defined by

\begin{equation}
    \langle \mathbf{q_1,q_2} \rangle_{x,{\Delta T}}=  \int_{\Delta T}^{} \int_{V}^{} \mathbf{q}_{1}^{*}(\mathbf{x},t)\mathbf{W(x)}\mathbf{q}_{2}(\mathbf{x},t) \,dV \,dt,
\end{equation}
where $\textbf{W(x)}$ is a diagonal positive definite weight matrix,
spatio-temporal structures $\boldsymbol{\phi}(\textbf{x},t)$ are identified by the maximization of
\begin{equation}
    \lambda=\frac{E\{|\langle \textbf{q}(\textbf{x},t),\boldsymbol{\phi}(\textbf{x},t) \rangle_{\textbf{x},\Delta T}|^{2}|r\}}{\langle \boldsymbol{\phi}(\textbf{x},t),\boldsymbol{\phi}(\textbf{x},t) \rangle_{\textbf{x},\Delta T}},
\end{equation}
using a variation approach. $\boldsymbol{\phi}(\textbf{x},t)$ and $\lambda$ are found from solving the following Fredholm eigenvalue problem

\begin{equation}
     \int_{\Delta T}^{} \int_{V}^{} \mathbf{C}(\mathbf{x,x'},t,t')\mathbf{W(x')}\boldsymbol{\phi} (x',t') \,dx' \,dt'=\lambda \boldsymbol{\phi}(\textbf{x},t),
\end{equation}
where $\textbf{C(x,x'},t',t)$ is the two-point space-time correlation tensor.
In discrete systems, $\boldsymbol{\phi}$ is obtained by an eigenvalue decomposition formulated on the two-point correlation tensor  
\begin{equation}
\textbf{Q}\textbf{Q}^{*}\textbf{W}\boldsymbol{\Phi}=\boldsymbol{\Phi}\boldsymbol{\Lambda},
\end{equation}
where $\textbf{Q}$ is the CPOD event realization matrix, $\boldsymbol{\Phi}$ contains eigenvectors $\phi_{i}(\textbf{x},t)$, and $\boldsymbol{\Lambda}$ is a diagonal matrix containing eigenvalues $\lambda_{i}$.
The eigenvectors are the CPOD modes and the eigenvalues are the corresponding energies.

The CPOD event matrix $\textbf{Q}$ differs from the standard snapshot POD input matrix.
Space-only POD requires an input data matrix $\textbf{Q}'$ with each column $i'$ containing flow-field information of a single snapshot $\textbf{q}_{i'}(\textbf{x)}$ over the entire statistically stationary time series, $i'=1,2,...,N$.
In contrast, CPOD flattens flow-field information from a consecutive sequence of snapshots, $\textbf{q}_{r}(\textbf{x}, t)$, into each column of $\textbf{Q}$ for every event realization $r=1,2,...,N_{R}$, where $N_{R}<<N$.
The CPOD sequence window $\Delta T$ (defined by the time preceding $t^{-}$ and following $t^{+}$ each event) is a parameter adjustable by the number of snapshots included in each column of $\textbf{Q}$.
Once the conditional filtering has selected $N_{R}$ realizations of each event, the data matrix $\textbf{Q}$ is populated as follows
\begin{equation}\label{eqn:Q_mat}
    \textbf{Q}=
    \begin{bmatrix}
q(t_{o}^{(1)}-t^{-}) & q(t_{o}^{(2)}-t^{-}) & 
\cdots & q(t_{o}^{(N_{r})}-t^{-})\\
\vdots & \vdots & 
\ddots & \vdots\\
q(t_{o}^{(1)}) & q(t_{o}^{(2)})  & \cdots & q(t_{o}^{(N_{r})})\\
\vdots & \vdots & \ddots & \vdots \\
q(t_{o}^{(1)}+t^{+}) & q(t_{o}^{(2)}+t^{+}) & \cdots & q(t_{o}^{(N_{r})}+t^{+})

\end{bmatrix}.
\end{equation}


In practice, CPOD modes can be calculated by a cheaper and faster singular value decomposition
\begin{equation}
\textbf{Q}=\boldsymbol{\Phi
\Sigma\psi^{T}},
\end{equation}
yielding $N_{R}$ CPOD modes $\phi_{i}(\textbf{x},t)$ with corresponding singular values ($\sigma_{1}$, $\sigma_{2}$, ... $\sigma_{N_{r}}$) in the diagonal matrix $\boldsymbol{\Sigma}$. 
The CPOD modes and singular values can be normalized and ranked by variance, similar to standard POD.
In addition to being a function of both space and time, CPOD modes are also orthogonal in the space–time inner product.



\subsection{Demonstrative example}

\subsubsection{Lorenz system}

\begin{figure}
\centering
\includegraphics[width=1\textwidth]{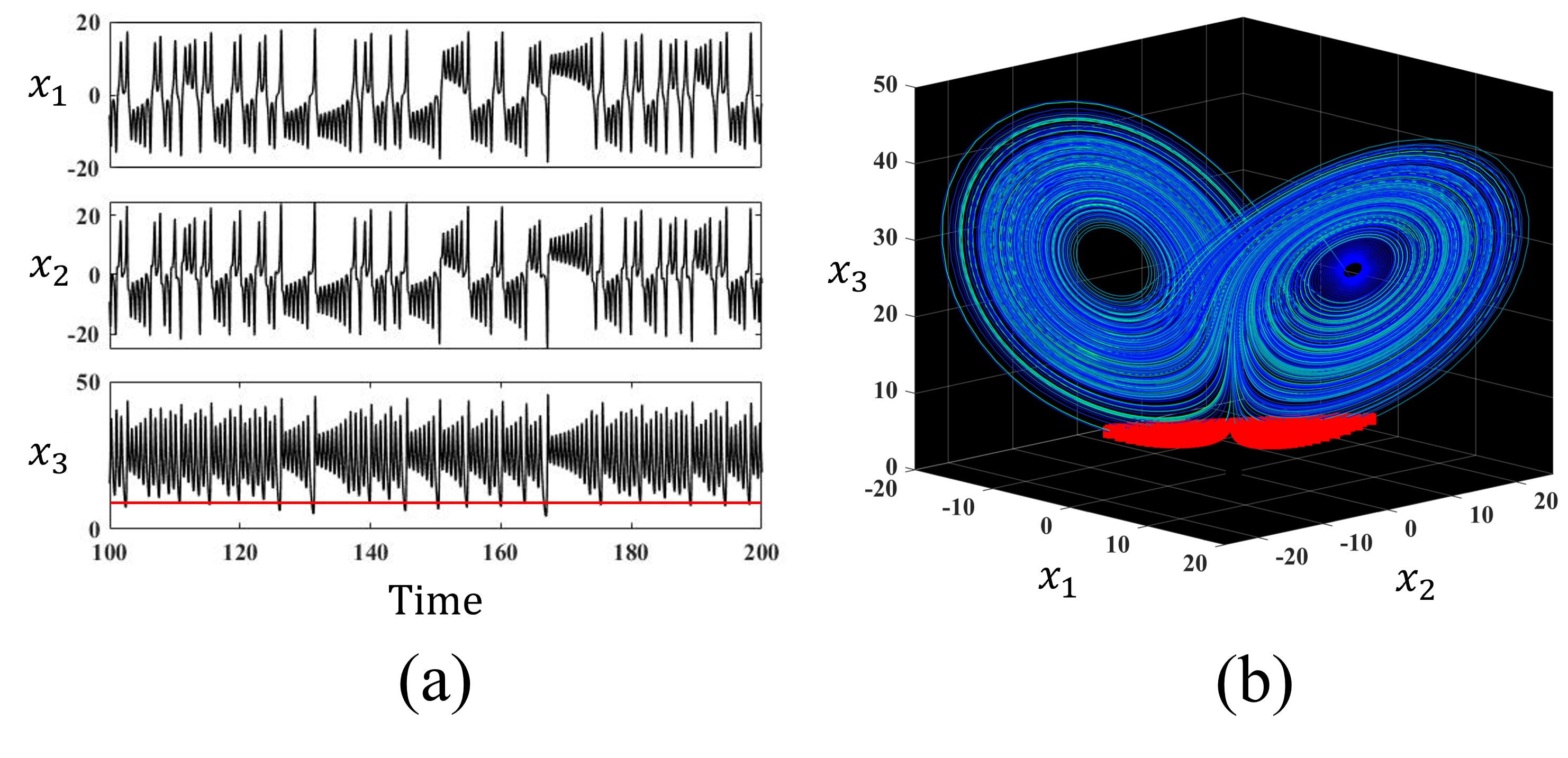}
\caption{Chaotic Lorenz system plotted as a function of (a) time and (b) phase-space. The CPOD event threshold (red line in (a)) is used to identify extreme fluctuations in $x_3$ that signify mode switching in $x_1$ and $x_2$. This corresponds to trajectories passing through the red region in (b) that switch attractor orbits.}
\label{fig:Lorenz}
\end{figure}

The CPOD methodology is now exemplified and compared to the closely related ensemble average.
Ideas as far back as Ruelle and Takens have suggested that turbulence is a manifestation in physical space of a strange attractor in phase space \cite{Ruelle1971}. 
In this spirit, CPOD is first instantiated on the classic Lorenz system \cite{lorenz1963deterministic,Brunton2017_rare}, which exhibits chaotic, intermittent behavior.
The non-linear three dimensional system, comprised of state-space observables $x_1, x_2$, and $x_3$, is presented in Fig.~\ref{fig:Lorenz}(a) as a function of time and (b) phase-space.
The system undergoes a bifurcation and oscillates around two attractors in an unpredictable manner. 
The inflection points from one attractor to another, are noticeable in $x_1$ and $x_2$ and are signified with a corresponding extreme fluctuation in $x_3$.
As such, $x_3$ is used to identify these switching dynamics as CPOD events, which are defined by any extreme negative fluctuation that crosses a threshold of $-2\sigma$ standard deviations of the signal (red line Fig.~\ref{fig:Lorenz}(a)).
In phase space, these events correspond to any trajectory along the manifold that passes through the red highlighted region in Fig.~\ref{fig:Lorenz}(b), thereby including all paths that switch orbits from either direction.

\begin{figure}
\centering
\includegraphics[width=1\textwidth]{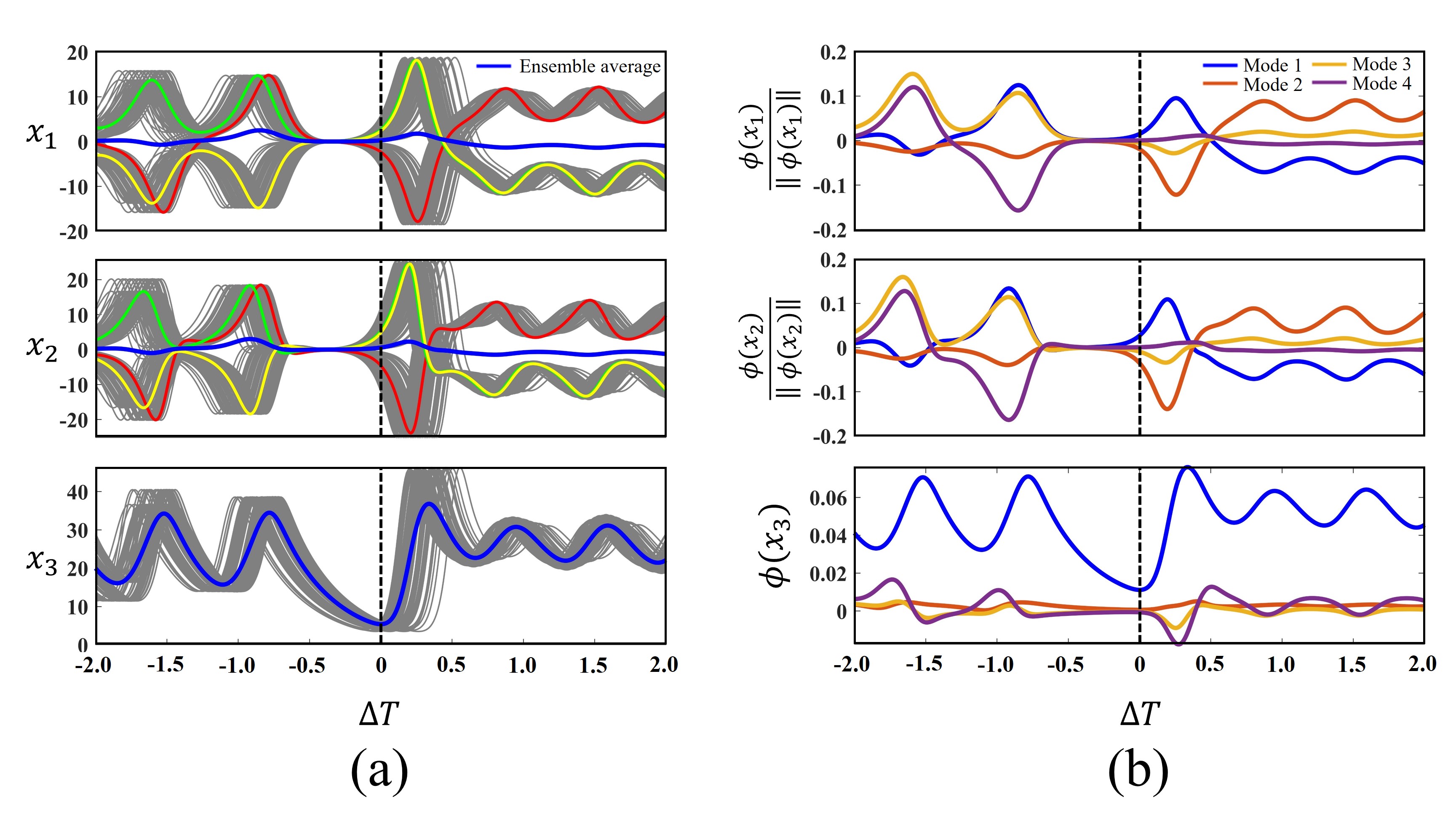}
\caption{State space variables of, (a) all CPOD events (grey) highlighting a few distinct trajectories (colored), and the ensemble average of all events (blue). Event sequences are identified by the extreme minimum of $x_3$ at time $t=0$ (dashed line).  (b) First four CPOD modes of the corresponding variables. $x_1$ and $x_2$ mode subsets are normalized to better distinguish mode trajectories. }
\label{fig:lor_events}
\end{figure}

The conditionally selected event trajectories are presented in Fig.~\ref{fig:lor_events}(a) for each state space variable.
Here, the time window $\Delta T= 4$ is centered around the event at local time $t=0$, with a total of $933$ event time instances collected (of $200,000$ samples); comments on their statistical independence is discussed later.
Examining $x_1$ and $x_2$ for the events, several distinct trajectory paths exists (a few examples are colored), inflecting up or down just before $t=0$.
Contrarily, $x_3$ only shows one characteristic trajectory with the extreme minimum fluctuation at the moment of the event. 
The ensemble average of all events is also shown Fig.~\ref{fig:lor_events}(a) with the blue curve.
In $x_1$ and $x_2$, the disadvantage of the ensemble average is clear; the average path does not accurately capture all the physical trajectories when multiple dynamic modes exist.
However, in the case of $x_3$, the ensemble average is able to represent the event, because the event itself is defined with respect to~$x_3$.
Taken together, the ensemble average intersects the phase-space manifold, but does not exist on it, thus it may not always faithfully reproduce the entire dynamic system.
In turbulence, this flaw is why Berkooz \textit{et al}. referred to the conditional sampling average of coherent structures as an art \cite{Berkooz1993}.

CPOD alleviates this shortcoming of the ensemble average by separating possible trajectories across a set of modes ranked by variance- or the expected probability of each event type. 
All three state variables are used in the CPOD event matrix $\textbf{Q}$, with the resulting first $4$ modes $\boldsymbol{\phi}_{i}(x_1,x_2,x_3,t)$ plotted in Fig.~\ref{fig:lor_events}(b). 
In each variable, the first CPOD mode directly reproduces the ensemble average; this connection will be elaborated on shortly.
Regarding the non-leading CPOD modes in $x_1$ and $x_2$, the different trajectory paths are now isolated and correlated to the other variables.
For plotting purposes, the $x_1$ and $x_2$ subsets of each CPOD mode are individually normalized by $||{\phi}(x)||$.
This normalization is omitted in $x_3$ to demonstrate the drop off in magnitude of non-optimal modes.
Alternatively, modes scaled by their singular values $\sigma_{i}$ could be a viable option to determine relative importance.
The corresponding singular value energy spectra will be examined following comments on the ensemble average.

Finally, comments are made on the conditional criteria used to identify the CPOD events. 
Here, the strategy of using a standard-deviation threshold of a signal to identify extreme events is straightforward.
However, if all time instances meeting this criteria are used, there may be more events selected than there are independent realizations.
This has the effect of oversampling some events because larger spikes linger below the threshold for a finite time, and therefore have multiple inputs into the CPOD data matrix.
Independent realizations are achieved here by a peak sampling algorithm that ensures no two events in the signal are within the same CPOD time window $\Delta T$, drastically reducing the number of realizations.
This more stringent conditional criteria may not always be necessary, as changes to results were insignificant for the Lorenz system.
In fact, the effects of oversampling could be beneficial in the sense of "smoothing the dynamics" \cite{Arbabi2017}; this is specifically addressed for fluid systems in section~\ref{secn:CPOD_int}.

\subsubsection{Connection to the ensemble average}

\begin{figure}
\centering
\includegraphics[width=1\textwidth]{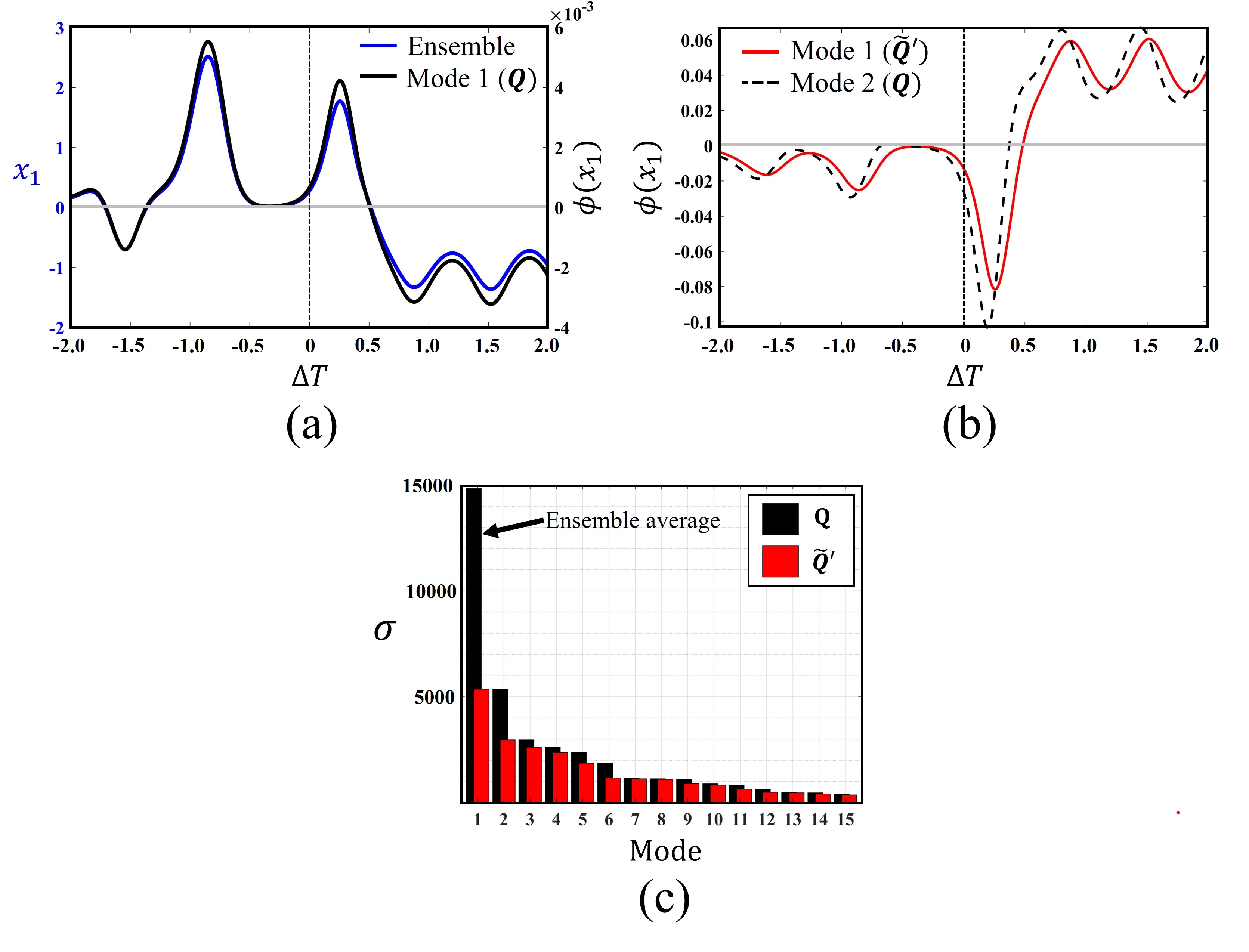}
\caption{(a) Comparison of the first CPOD mode ($x_1$) using whole quantities $\textbf{Q}$ and the ensemble average $\boldsymbol{\tilde{Q}}$. (b)~CPOD modes of the ensemble average subtracted data $\boldsymbol{\tilde{Q}}'$ reproduce the subsequent CPOD modes of $\textbf{Q}$. (c)~Singular values of both data types share this pattern and are dominated by the first mode. }
\label{fig:lor_inputs}
\end{figure}

The connection between the ensemble average and CPOD is discussed by first considering space-only POD. 
Typically, POD input data consists of fluctuating quantities $\textbf{q}'(t)$ around the time averaged mean $\boldsymbol{\bar q}$ defined by 
\begin{equation}
    \boldsymbol{\bar q} = \frac{1}{N}\sum_{n=1}^{N}\textbf{q}(t).
\end{equation}
It is well established that if the whole quantity $\boldsymbol{q}(t) = \boldsymbol{\bar q} + \boldsymbol{q'}(t)$ is used instead of fluctuations, the first POD mode will reproduce the time-mean, as POD is optimized for the mean square of the input variable. 
By analogy, the situation for space-time CPOD also necessitates examining the choice of input variable.
Here the CPOD input is the entire quantity $\textbf{Q}(t,r)$, and the time-mean is replaced with the ensemble average of event sequences $\boldsymbol{\tilde{Q}}(t)$ , defined by
\begin{equation}
        \boldsymbol{\tilde Q}(t) = \frac{1}{N_R}\sum_{r=1}^{N_R}\boldsymbol{Q}(t,r).
\end{equation}
Fluctuations about the ensemble average $\boldsymbol{\tilde{Q}}'(t,r)$ can similarly be decomposed as $\boldsymbol{Q}(t,r) = \boldsymbol{\tilde{Q}}(t) + \boldsymbol{\tilde{Q}}'(t,r)$ and used as CPOD input.
This is demonstrated with the Lorenz example, which originally used the whole quantity $\textbf{Q}(t,r)$, and is now recalculated with $\boldsymbol{\tilde{Q}}'(t,r)$.   
Figure~\ref{fig:lor_inputs} plots the first modes of each case and compares with the ensemble average for the state space variable $x_1$.
In Fig.~\ref{fig:lor_inputs}(a), it is confirmed that the first CPOD mode of data-type $\textbf{Q}$ is the ensemble average, but scaled.
Analogous to POD, Fig.~\ref{fig:lor_inputs}(b), shows that the first CPOD mode of  $\boldsymbol{\tilde{Q}}'$ is nearly identical to the the second mode of $\textbf{Q}$.
This pattern continues as ${\phi_{i}}(\boldsymbol{\tilde{Q}}')={\phi_{i+1}}(\boldsymbol{Q})$ for all following modes.
Figure~\ref{fig:lor_inputs}(c) reveals the same pattern exists for the singular values ${\sigma_{i}}(\boldsymbol{\tilde{Q}}')={\sigma_{i+1}}(\boldsymbol{Q})$.
The first mode dominates in each case, but is substantially larger when the ensemble average is included ($\textbf{Q}$), which is also typical for the fluid systems encountered in this work.

Despite the analogous relationship to POD, which usually subtracts the time-mean component, it is concluded that the ensemble average should be included in the CPOD analysis.
This is because the ensemble average (or equivalently CPOD mode 1 of $\textbf{Q}$) likely contains the most relevant dynamics sought, as demonstrated in $x_3$ of the Lorenz system.
Of course, the interpretation of the CPOD modes should be grounded in the context of the known physics, particularly when determining the cutoff of relevant modes.
Typically only the first, or first two modes are chosen for the remainder of the paper.

\section{CPOD for flow event characterization }\label{secn:CPOD_int}

\subsection{Supersonic boundary layer transition}
\subsubsection{Intermittent CPOD events}
\begin{figure}
\centering
{\includegraphics[trim=0 0 0 0,clip,width=1.0\textwidth]{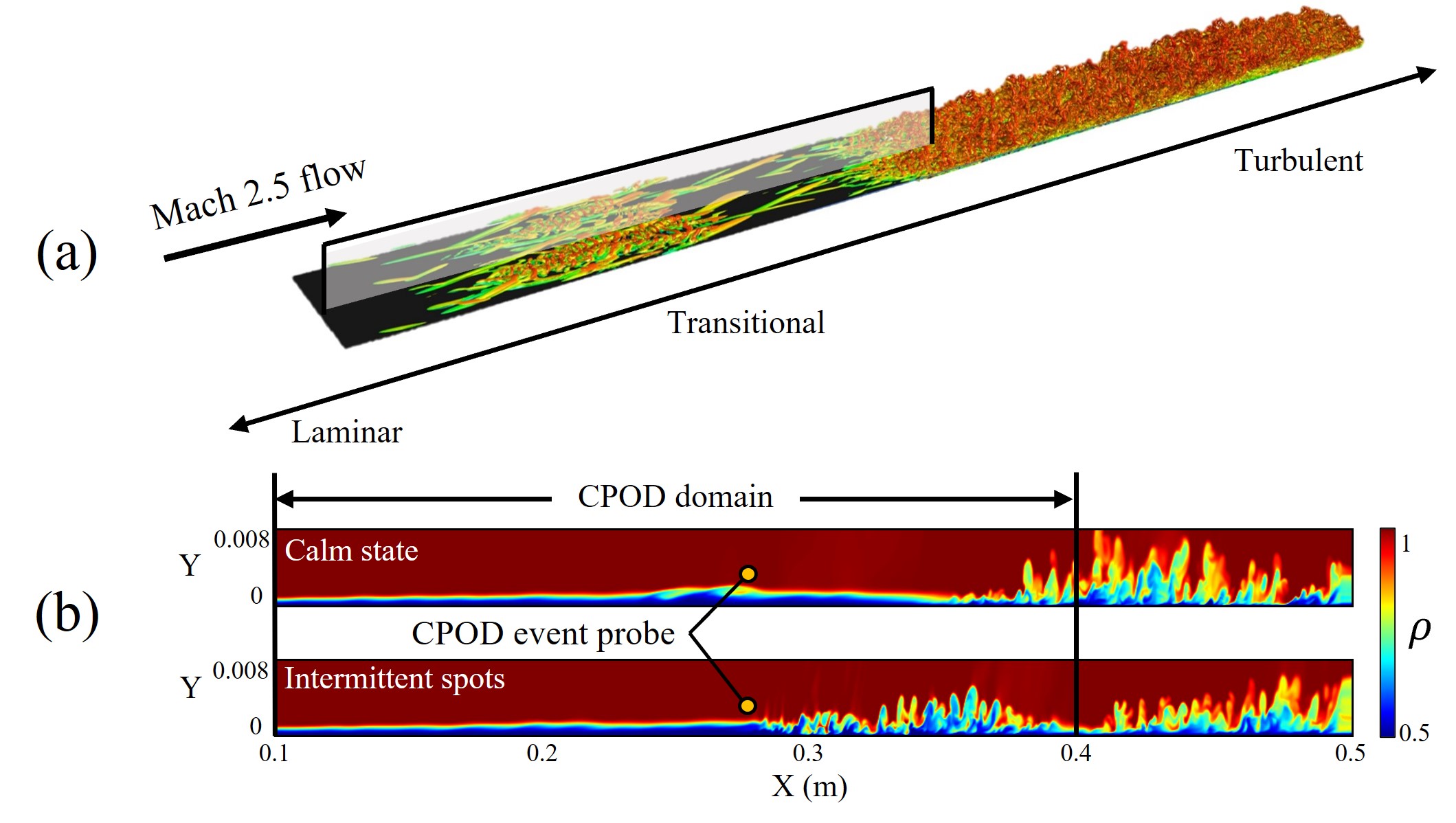}}
\caption{(a) Vorticity isosurfaces~$(Q=100)$ of a Mach~$2.5$ flat plate, transitional boundary layer. (b) Laminar-to-turbulent transitional regime undergoes intermittent turbulent spots. A density probe (yellow circle) is placed at the initial instability location to characterize the CPOD events.
The sub-domain for the CPOD analysis is marked in (b). }
\label{fig:flat_instant}
\end{figure}

\begin{figure}
\centering
\includegraphics[width=1\textwidth]{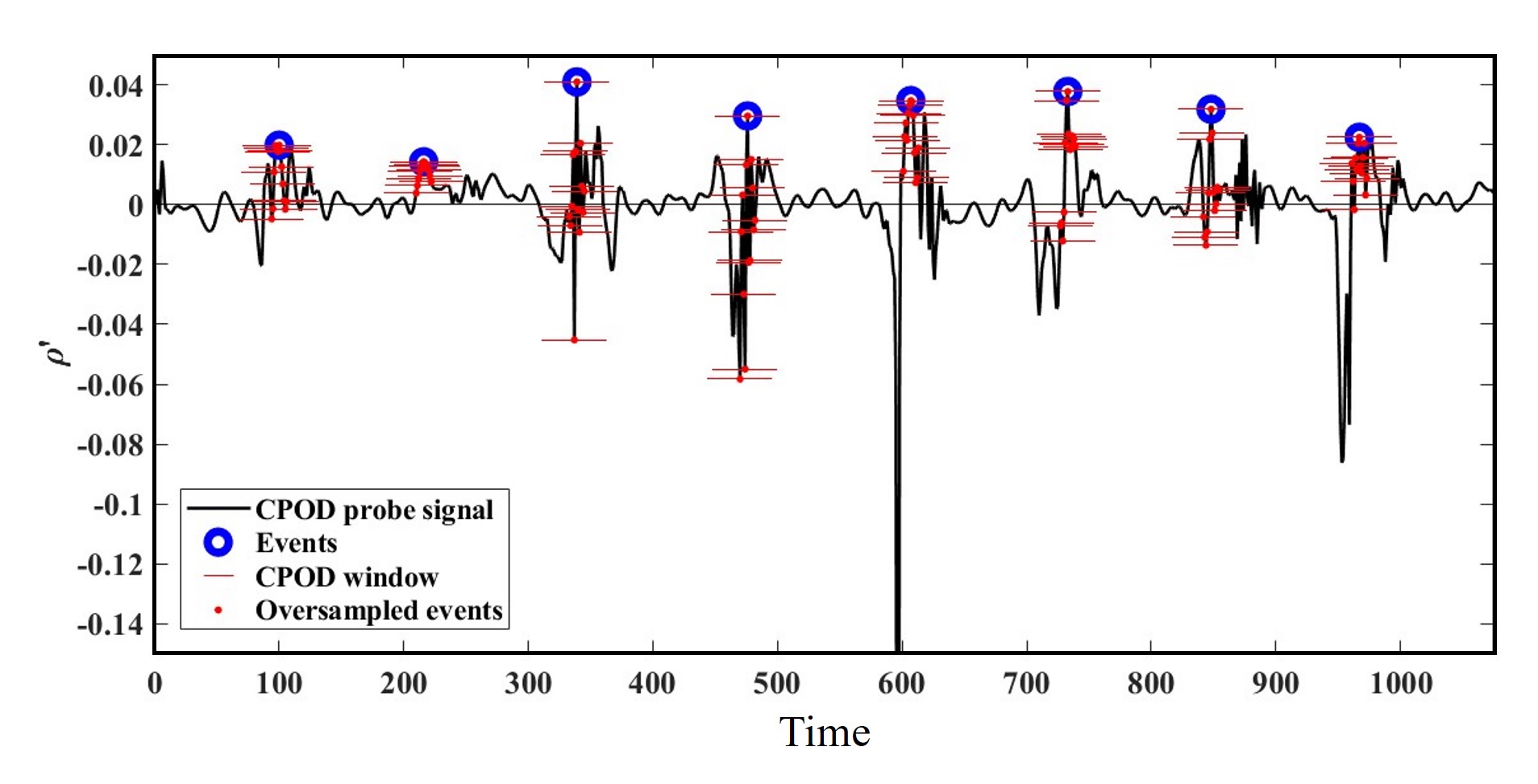}
\caption{CPOD density probe signal identifies the peak of intermittent turbulent events (blue circles). The $13$~closest time instances (red circles) test the effects of oversampling on CPOD modes. Horizontal lines span the CPOD event time-window. }
\label{fig:flat_sig}
\end{figure}

The basic elements of a CPOD analysis applied to a complex flow are now demonstrated before extending the method in later sections. 
The system under consideration is the numerical, laminar-turbulent transition of a flat plate boundary-layer at Mach~$2.5$, introduced in Fig.~\ref{fig:flat_instant}(a).
Details regarding the problem set up are found in~\citet{goparaju2022supersonic}.
Here, the transitional features in the cross-(symmetry) plane are illustrated through the density field in Fig.~\ref{fig:flat_instant}(b) at two representative time instances.
The phenomenon to be elucidated with CPOD is the intermittent high-frequency structures resembling turbulent spots in the late transitional regime.
The seemingly random appearance of these fluctuations are a consequence of the stochastic distribution of freestream disturbances~\cite{shaikh1997investigation}, and are realized here through receptivity forcing informed by wind-tunnel broadband acoustic disturbances \cite{Duan2019,goparaju2022supersonic}.
Resulting intermittent spots are accompanied by boundary-layer acoustic radiation~\cite{wang1996sound} and enhanced wall friction~\cite{marxen2019turbulence}, the characterization of which is vital for high-speed vehicle design. 
In order to specify these processes as a conditional event for the CPOD algorithm, a probe is placed in the freestream as labeled in Fig.~\ref{fig:flat_instant}(b).
The intermittent nature is further exhibited by the extreme density fluctuations of the CPOD probe signal in Fig.~\ref{fig:flat_sig}.

Several aspects of the CPOD analysis pertaining to event selection are discussed using the signal of Fig.~\ref{fig:flat_sig}.
Firstly, events must be conditionally defined; here, $8$~local peaks (blue circles) in the density signal were manually identified at the center of each burst.
Secondly, a CPOD time-window period $\Delta T$ is set; here $60$~snapshots are chosen across each event sequence (spanned by red bars).
For the current purpose, this window provides material insights into the intermittent nature of the problem, as discussed shortly.
The role of event oversampling is also investigated in the extreme with a second CPOD analysis.
The oversampled case, also represented in the signal of Fig.~\ref{fig:flat_sig}, encompasses $104$~event realizations (red circles) by selecting the $13$~nearest time-instances for each unique event.
The value of this analysis is in understanding the effects incurred by other conditional event selection methods, such as the standard deviation threshold functions used in the Lorenz system and throughout the paper, as well as in similar instances elsewhere~(e,g. \citet{hack2021extreme}).

\begin{figure}
\centering
\includegraphics[width=1\textwidth]{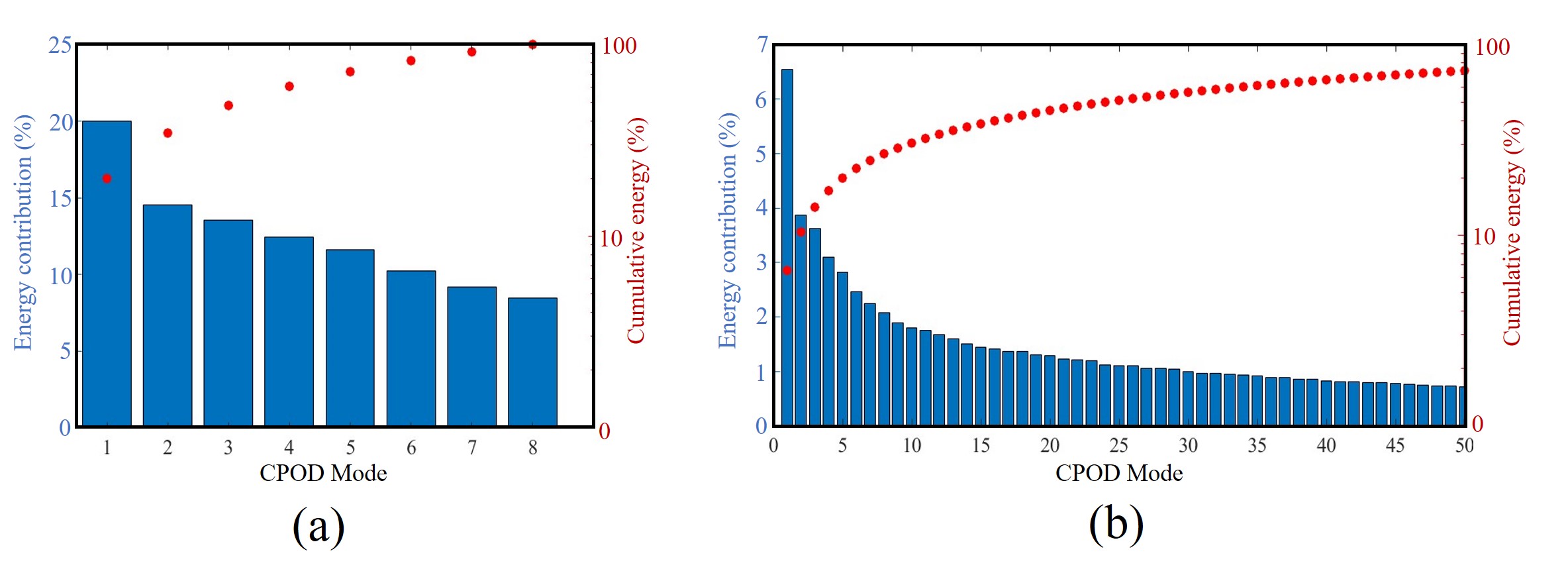}
\caption{CPOD singular values as a percentage of mode energy contribution for (a)~the $8$ unique events and (b)~oversampled event realizations.}
\label{fig:flat_singlulars}
\end{figure}

The CPOD analyses are conducted on density fluctuations in the transitional regime spanning from $X=0.1m$ to $0.4m$. 
The resulting CPOD singular values are examined in Figs.~\ref{fig:flat_singlulars}(a,b) for the $8$~unique events, and the oversampled case, respectively.
The singular values are expressed as a percentage of energy contributed from each mode.
The largest difference in the modal energy is between the first and second modes, irrespective of the sampling rate, which alludes to low-rank behavior.
Higher modes (second and above) have lower energy differences, as indicated by the flattening of cumulative energy distribution in Fig.~\ref{fig:flat_singlulars}. 
The effect of oversampling in (b), produces a more prominent leading mode than (a), nearly doubling the contribution of the first mode relative to the second. 
This surprising augmentation of the leading mode is beneficial in the sense of reduced-order representation, and its implications are examined in the next subsection.

\begin{figure}
\centering
\includegraphics[width=1\textwidth]{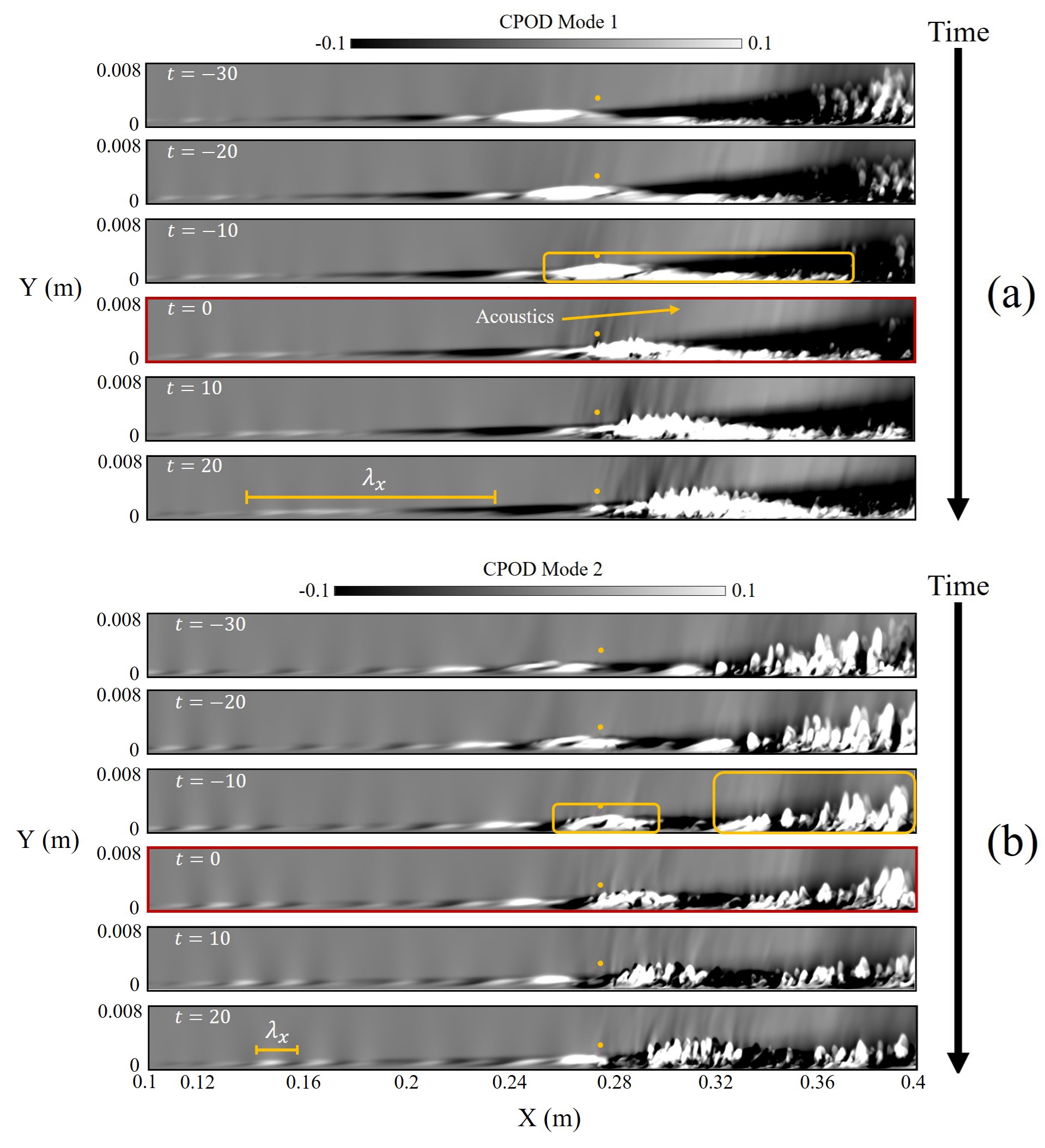}
\caption{CPOD mode sequences of the intermittent turbulent spot progressing in time from top to bottom. (a)~Mode~$1$ shows an elongated coherent structure (yellow box) and the formation of acoustic waves generated at the CPOD probe signal (yellow circle) at $t=0$. (b)~Mode~$2$ displays a second event type consisting of smaller structures generated in quick succession.   }
\label{fig:flat_modes}
\end{figure}

The CPOD mode structures corresponding to the $8$~unique events are first presented. 
In Fig.~\ref{fig:flat_modes} the two leading CPOD modes are shown.
The modes progress in time from top to bottom, with the peak event density fluctuations occurring in the red frame at~$t=0$.
As noted earlier in the Lorenz system, the first CPOD mode has a clear definition as the ensemble average.
Examining CPOD mode~$1$ at this moment, a coherent structure passes under the probe (yellow dot) at this instance.
Strong freestream density fluctuations (acoustic waves) are also observed, oriented obliquely and traveling downstream as the flow progresses.
The intensity of the acoustic radiation is lower in mode~$2$, consistent with the energy distribution in Fig.~\ref{fig:flat_singlulars}(a).
The boundary-layer structures upstream of the intermittent spot~($X<0.24m$), exhibit longer wavelengths ($\lambda_x$) in mode~$1$ compared to mode~$2$.
Related, CPOD mode~$1$ demonstrates an event type consisting of a single elongated structure (yellow box), as opposed to mode~$2$ which indicates smaller turbulent spots that follow each other in rapid succession.

\subsubsection{Oversampling}
\begin{figure}
\centering
\includegraphics[width=1\textwidth]{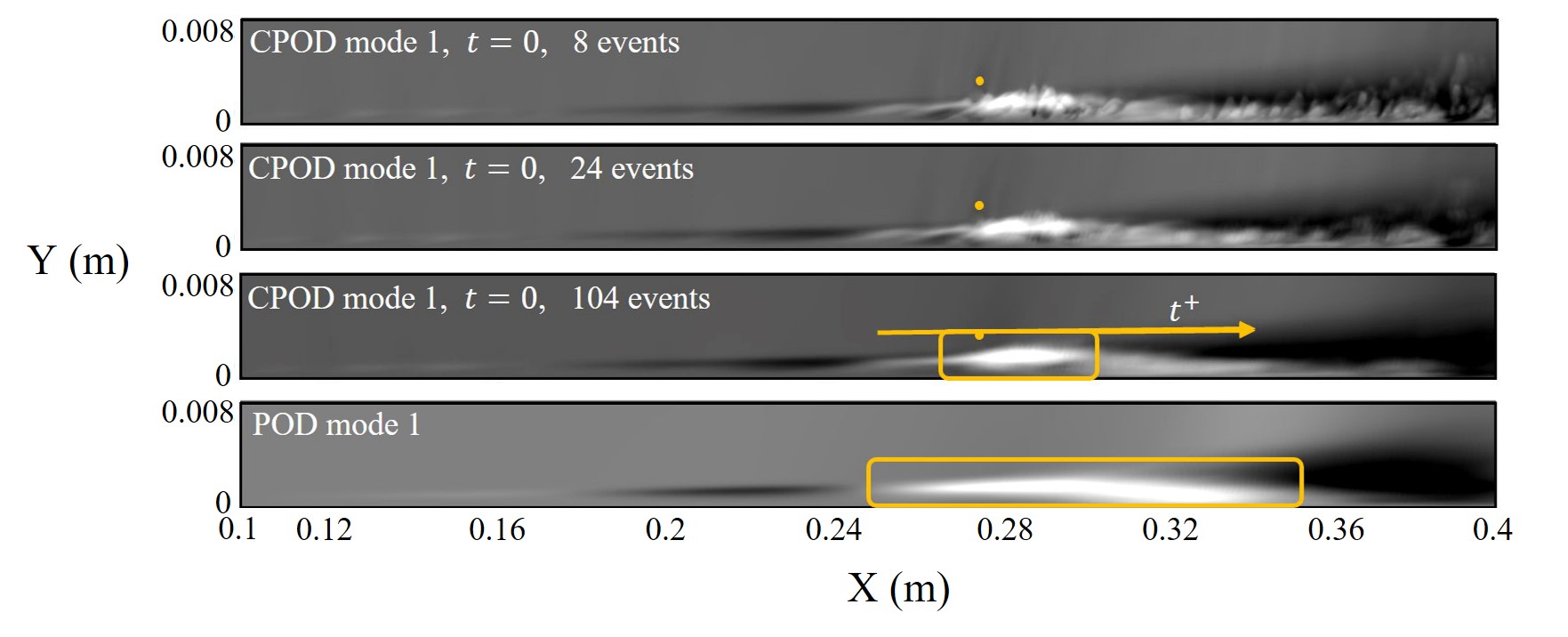}
\caption{CPOD mode~$1$ snapshot comparing the effects of progressively oversampled events. Colors are scaled by the minimum and maximum of each mode for a relative comparison. Coherent structures become averaged-out and qualitative similarities to the leading mean-subtracted POD mode become apparent.  }
\label{fig:flat_modes_compare}
\end{figure}

The effects of oversampled events on the CPOD results are now discussed. 
For this, the entirety of the CPOD mode sequence is not necessary, and only the frame at the moment of the event ($t=0$) is presented in Fig.~\ref{fig:flat_modes_compare} for $8, 24,$ and $104$ CPOD event realizations.
Note that in this figure, the gray contours scale with the minimum and maximum of the individual mode amplitude, whereas Fig.~\ref{fig:flat_modes} was clipped to clearly illustrate the smaller amplitude acoustic waves.
Here, it is apparent that progressive oversampling averages out the smaller fluctuations and smooths the evolution of coherent structures.
Similar smoothing behavior was also found for the oversampled CPOD mode~$2$, which maintained its smaller wavelength structures and is omitted for redundancy.
In this manner, an oversampled conditional event criteria could be advantageous in terms of further reducing the complexity of the flow, and thus accounting for the magnification of the leading CPOD mode singular value in Fig.~\ref{fig:flat_singlulars}(b).

In some ways, the oversampled CPOD results qualitatively approach the leading mean-subtracted POD spatial mode, shown for reference in Fig.~\ref{fig:flat_modes_compare}.
However, there are several differences inherent to each method that should be reiterated in context of these results. 
Principally, CPOD is time-local, whereas POD is not coherent over time, and includes data from every snapshot.
Thus we see how the peak correlation (yellow box) in the CPOD mode is localized, and would travel from left-to-right over the course of the CPOD trajectory. 
Contrarily, POD spreads the peak correlation over the entire length of the turbulent spot, losing temporal resolution. 
There is also a major difference in the behavior of non-leading POD modes, where even qualitative comparisons with CPOD modes break down.
For example, in Fig.~\ref{fig:flat_modes}(b), the smaller wave-length structures preceding the turbulent spot in CPOD mode~$2$ only become realized in POD after the tenth mode.
These discrepancies result in the intermittent dynamics being spread across a much larger set of POD modes, where for context, the first and second modes only capture $2$ and $1.8\%$ of the energy, respectively. 
The high-rank nature of POD modes is typical for other transitional boundary layers~\cite{hosseinverdi2019numerical}, highlighting the indispensable role of the conditional space-time adaptation to extract more meaningful insights in this class of problems.

\begin{figure}
\centering
\includegraphics[width=1\textwidth]{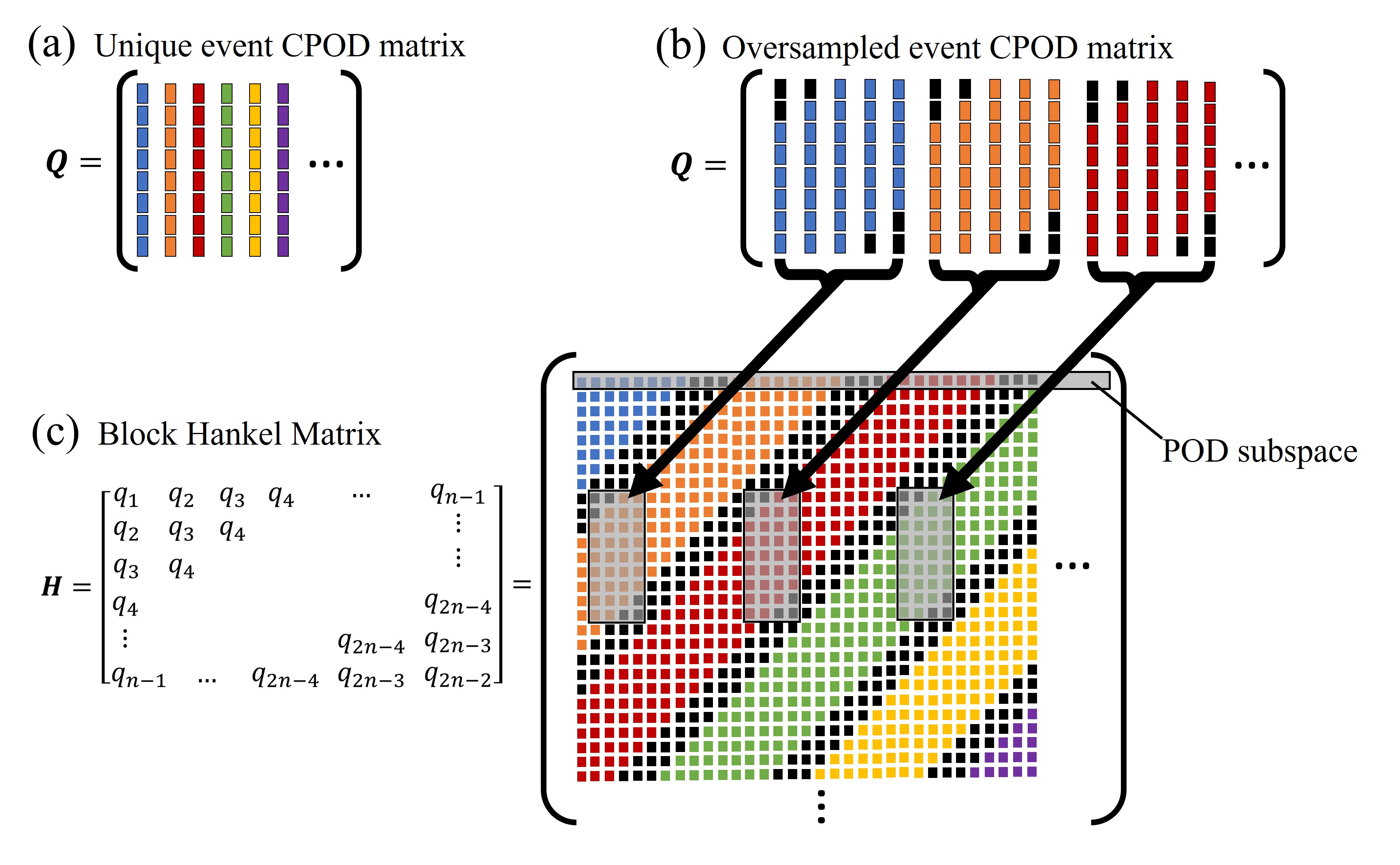}
\caption{Comparison of the (a) unique event CPOD data matrix, (b) oversampled CPOD matrix and (c) block-type Hankel matrix (c). The oversampled matrix structure is a subset of the larger Hankel space and benefits from the time-delay embedding properties. }
\label{fig:Q_matrix}
\end{figure}

The favorable properties of oversampling are expounded upon by examining the general structure of the CPOD event matrix $\textbf{Q}$.
In Fig.~\ref{fig:Q_matrix}(a), the block representation of snapshots, colored by their unique event sequence, are flattened into each column of the matrix following Eqn.~\ref{eqn:Q_mat}.
In the oversampled event matrix (b), the additional shifted sequences incorporate snapshots adjacent to the original event window (black blocks).
Insights into this time-shifted pattern are inferred by considering the structure of a block-type Hankel matrix (c). 
This data arrangement has been exploited, for example, in the shift-stacking of snapshot matrices to improve DMD results \cite{Tu2014}, the Eigensystem Realization Algorithm (ERA) \cite{Juang_1986,Ma_2011} and many more decompositions \cite{Schmid2021_review}.
These methods capitalize on the time-delay coordinates embedded in the shifted Hankel structure \cite{Takens1981}, which offers a mathematical connection to Koopman theory \cite{Arbabi2017} and superior observably of nonlinear intermittent dynamics \cite{Brunton2017_rare}. 
Recently, mathematical connections between the classical space-time POD correlation and Hankel matrices has been put forth \cite{Frame2022}, demonstrating the approximation of Hankel modes.
Here in Fig.~\ref{fig:Q_matrix}, we see how the oversampled Conditional space-time POD event matrix comprises time-local subsets of the Hankel matrix.
Thus, performing the SVD on the oversampled CPOD matrix benefits from partially capturing the time-delay embedding properties, while at a fraction of the cost of decomposing the massive Hankel matrix. 
For comparison, the subspace spanned by the POD snapshot matrix is limited to a single row of the Hankel matrix.

\subsection{Video processing: inlet unstart buzz}

\begin{figure}
\centering
\includegraphics[width=1\textwidth]{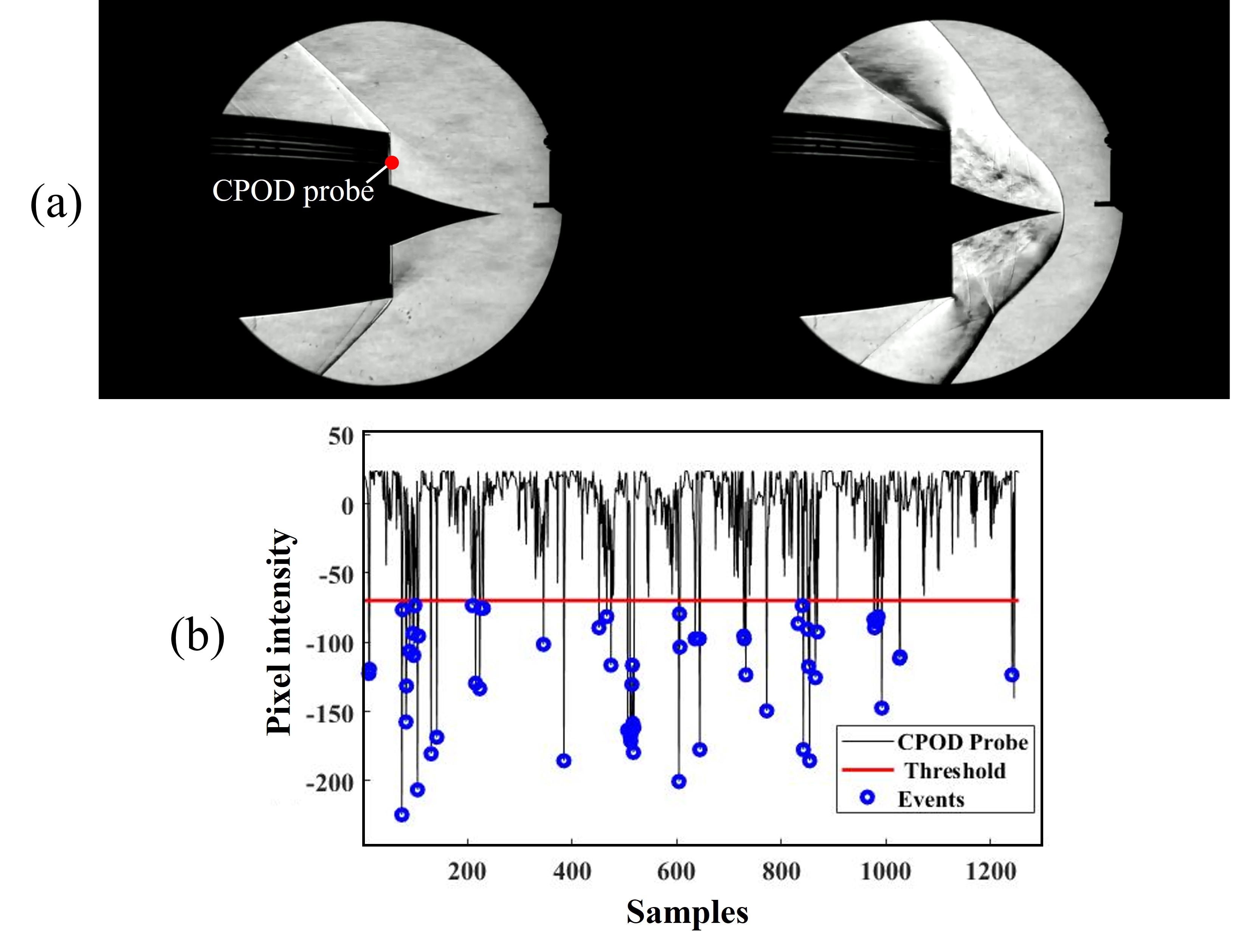}
\caption{(a) Schlieren images of Mach $1.7$ inlet buzz showing nominal and forward shock positions\cite{Chima2012}. (b) Pixel intensity probed at the inlet face (red dot) identifies the events that trigger the unstart process. }
\label{fig:buzz}
\end{figure}

CPOD is also applicable to experimental data such as the video processing of Schlieren imaging or Particle Image Velocimetry (PIV).
As a brief example, CPOD is used to analyze pixels from a Schlieren video of Mach $1.7$ flow undergoing engine inlet "buzz" with extreme shock oscillations along a compression spike \cite{Chima2012}.
The configuration and shock motion is introduced in Fig.~\ref{fig:buzz}(a), demonstrating the nominal and unstarted shock positions.
The unsteady conditions at the inlet face (red dot) are examined with CPOD by identifying intermittent fluctuations of pixel intensity, 
plotted in Fig.~\ref{fig:buzz}(b).
The event threshold function ($2\sigma$ standard deviations) identified $59$ extreme instances and uses a time window spanning $5$ to $10$ frames before and after each event, respectively.

\begin{figure}
\centering
\includegraphics[width=1\textwidth]{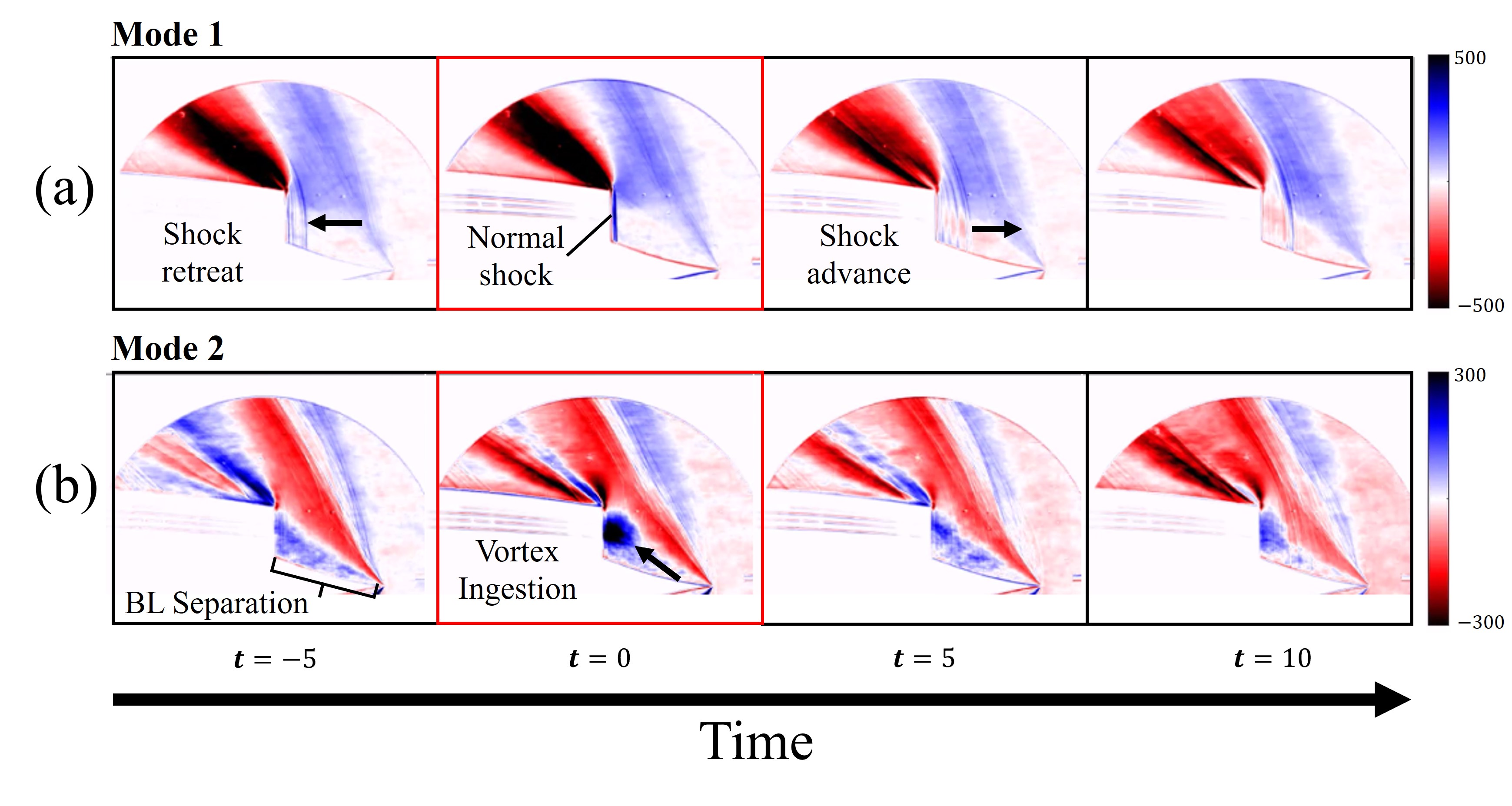}
\caption{CPOD modes of the unstart oscillation progressing in time from left to right. (a) Mode~$1$ is correlated to the forward and backward motion of the shock, which as normal to the inlet face at event time $t=0$. (b) Mode~$2$ shows a separated vortex from the boundary layer that is ingested at $t=0$, triggering the shock advancement.   }
\label{fig:buzz_CPOD}
\end{figure}

Figure~\ref{fig:buzz_CPOD} shows the first two CPOD modes progressing in time from left to right. 
Comparing the modes, CPOD clearly distinguishes two separate behaviors, relating the shock motion (mode $1$) and the vortex separated from the compression spike (mode $2$). 
Examining Fig.~\ref{fig:buzz_CPOD}(a), a shock wave is observed moving back and forth along the spike, matching the inlet face with a normal shock at the moment of the extreme fluctuation.
The onset of the buzz instability is intermittent, and lasts between 4 to 6 of these shock oscillations.
Throughout the relatively shorter CPOD sequence window, the boundary layer has already separated from the spike. 
As such, mode $2$ (Fig.~\ref{fig:buzz_CPOD}(b)) shows a large vortex structure about to pass through the normal shock and enter the engine at the event time instance, which induces the upstream shocks advancement at $t=10$ in mode 1.
This example demonstrates how CPOD can robustly separate low rank dynamics (12 and 5\% energy in the first and second modes) directly mapped from movies of experiments.



\section{CPOD and DMD for stability analysis}\label{secn:CPOD_DMD}

\begin{figure}
\centering
\includegraphics[width=1\textwidth]{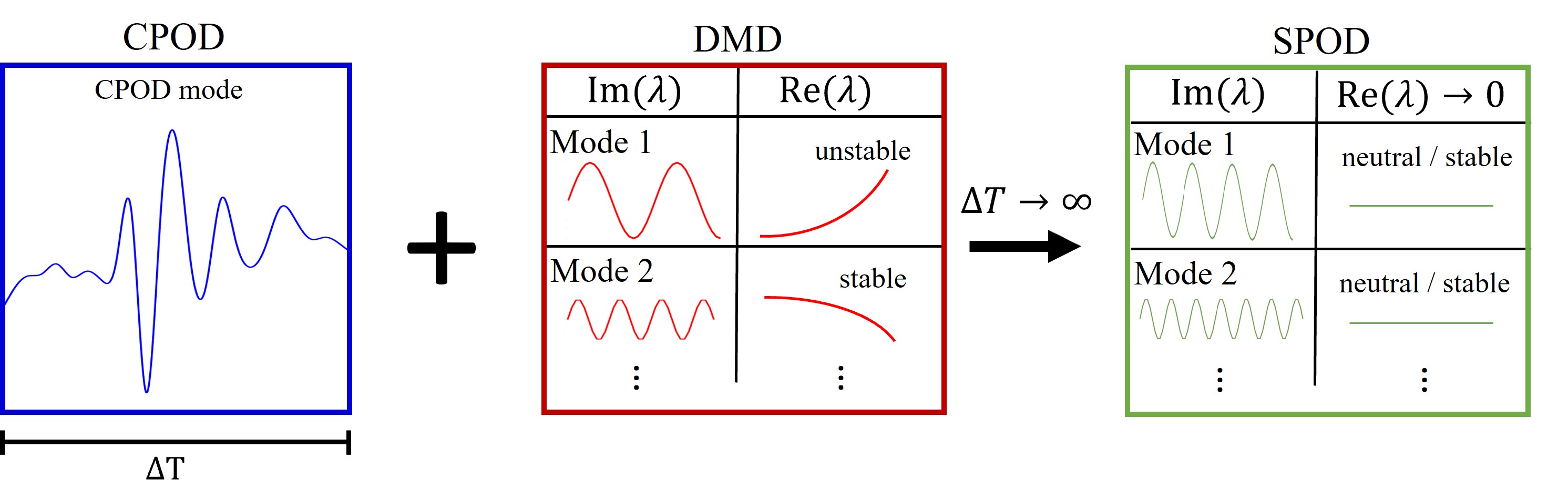}
\caption{Transient CPOD modes contain frequency and growth rate information that can be isolated in complex DMD eigenvalues $\lambda$. As the CPOD time window $\Delta T$ increases, growth rates become insignificant and CPOD-DMD modes begin to approximate SPOD modes.   }
\label{fig:mode_relations}
\end{figure}

In this section, CPOD is extended with a subsequent Dynamic Mode Decomposition and connections are made to spectral and stability analysis.
It will be shown that the transient CPOD modes contain information in a reduced subspace that DMD can use to extract frequency and growth rate information, as abstracted in Fig.~\ref{fig:mode_relations}.
Two methods are proposed based on this fact.
1) A CPOD-DMD analysis that produces tonal modes; this is proven to exactly recover leading SPOD modes if $\Delta T$ is sufficiently long enough to nullify the real part of the DMD eigenvalues.
2) A multi-resolution DMD framework (CPOD-mrDMD) that yields a "cause and effect" stability analysis by considering real eigenvalue growth rates that are significant over shorter time-horizons.

This section focuses on the implementation of these two CPOD extensions.
The DMD theory and algorithm applied to the CPOD modes is outlined in Appendix~\ref{secn:appen_DMD}, while mathematical relations between CPOD, SPOD, and DMD can be referenced elsewhere \cite{Towne2018_relation,Tu2014}.
Both CPOD-DMD extensions are exemplified with a resonating supersonic impinging jet.
The impinging jet is ideal for elucidating both global resonance tones generated from a self-sustained aeroacoustic feedback loop (CPOD-DMD in Section~\ref{secn:DMD_sub_DMD}), and the local convective instabilities resulting from the acoustic forcing process within the feedback loop (CPOD-mrDMD in Section~\ref{secn:DMD_sub_mrDMD}).

\subsection{CPOD of a resonating impinging jet}\label{secn:DMD_sub_SIJ}

\begin{figure}
\centering
\includegraphics[width=1\textwidth]{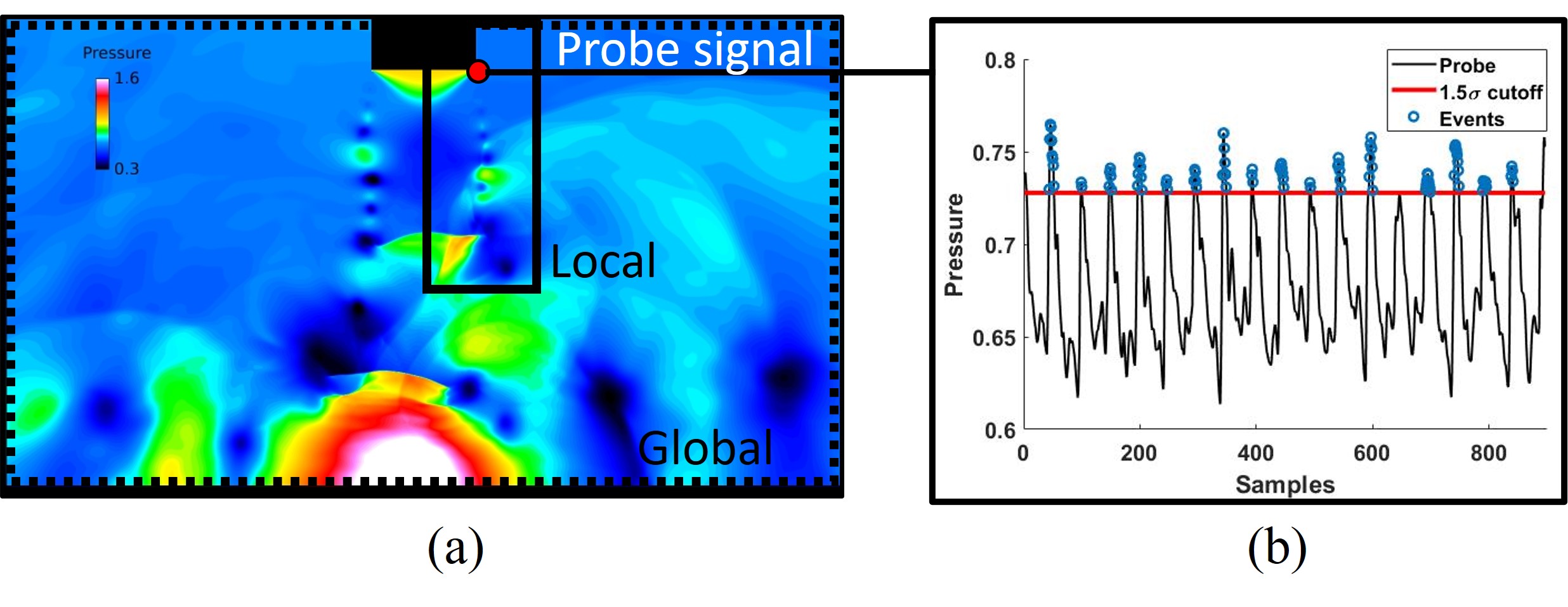}
\caption{Supersonic impinging jet demonstrating the acoustic feedback forcing. (a) Pressure flow-field showing the receptivity probe location, local and global domains used in the CPOD analysis. (b) The probe signal with a $1.5\sigma$ standard deviation threshold identifies high amplitude fluctuations associated with the periodic feedback forcing events.}
\label{fig:imping_intro}
\end{figure}

The initial CPOD analysis of the resonating impinging jet is first introduced before examining the DMD extensions.
Specifically, a LES of an underexpanded (NPR $2.65$), Mach $1$, planar jet impinging at a height of $4$ nozzle widths is studied.
An instantaneous snapshot of the pressure field is shown in Fig.~\ref{fig:imping_intro}(a) for reference. 
Further details of the numerical methods and configuration are reported in \cite{stahl_caes2022}.
The CPOD events in question pertain to the acoustic receptivity at the jet nozzle, where upward propagating acoustic waves perturb the initial shear-layer and trigger a Kelvin-Helmholtz instability. 
The resulting shear-layer growth produces structures that impinge on the ground plane, generating the next upstream propagating acoustic wave that closes the feedback loop.
Figure~\ref{fig:imping_intro}(b) captures the signature of the repeated acoustic forcing in the pressure signal probed at the nozzle receptivity location. 
Since the forcing events are periodic in nature, a simple $1.5\sigma$ standard deviation threshold function suffices to identify the acoustic receptivity events.

\begin{figure}
\centering
\includegraphics[width=.7\textwidth]{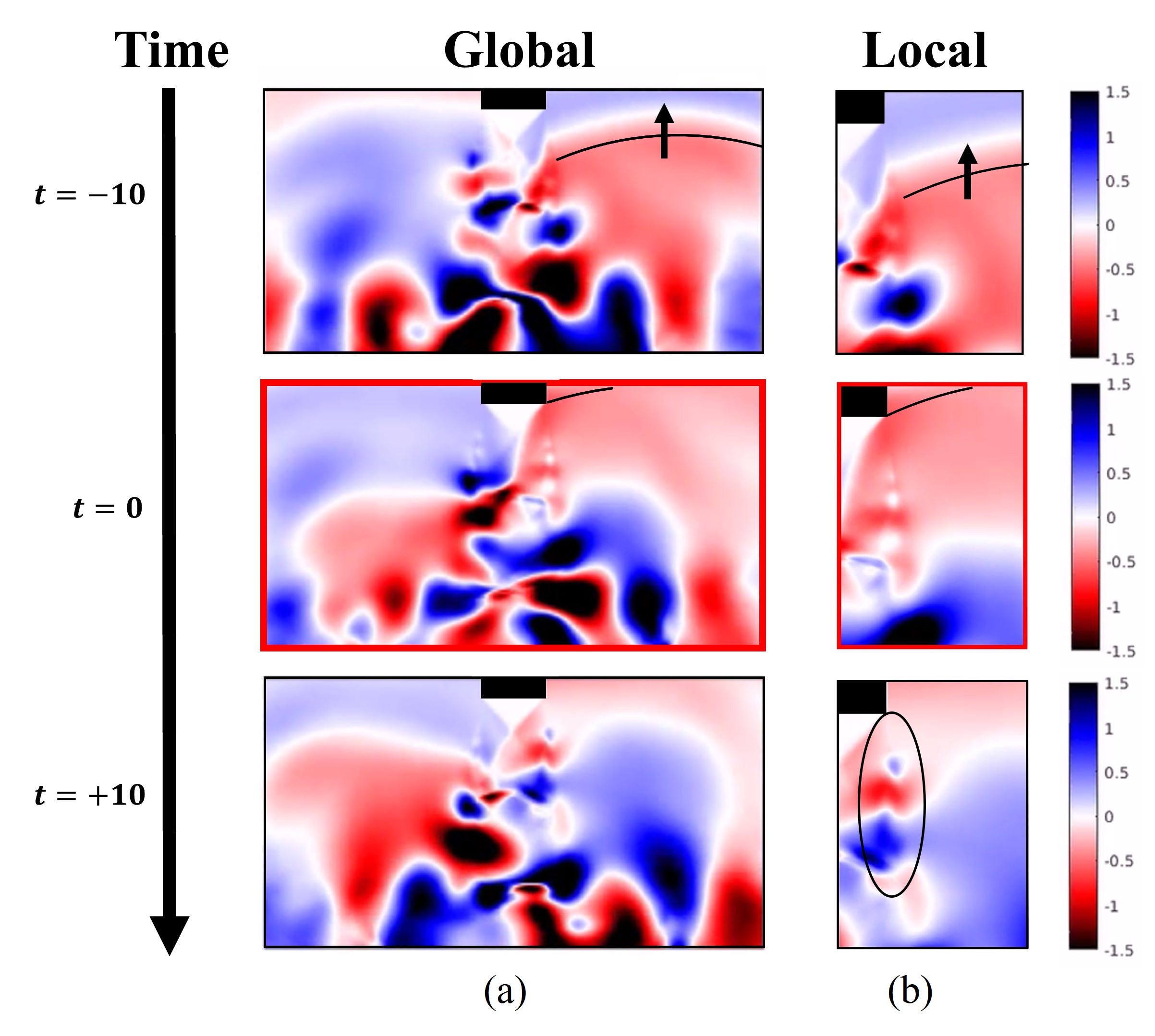}
\caption{Leading CPOD modes correlated to the acoustic receptivity forcing event of a supersonic impinging jet, progressing in time from top to bottom in the (a) global and (b) local domains. }
\label{fig:loc_glob}
\end{figure}

The corresponding pressure-field fluctuations are used to calculate two separate CPOD analyses on a global and local spatial domain, as demarcated in Fig.~\ref{fig:imping_intro}(a).
The global domain will be used for the CPOD-DMD spectral analysis and the local domain will be used for the CPOD-mrDMD stability analysis.
The resulting first CPOD modes are presented in Fig.~\ref{fig:loc_glob} for the (a)~global and (b)~local domains, progressing in time from top to bottom.
The first CPOD mode contains $14$ and $11\%$ of the total energy, for (a) and (b) respectively, compared to approximately $7\%$ in the second mode for both domains.
In each case, the acoustic wave is observed hitting the right side of the nozzle at time $t=0$, followed by wavepacket growth in the shear-layer.
A wealth of insight can be gained from the CPOD modes alone, as has been done previously \cite{stahl_sci2022}.

Anticipating the upcoming DMD results of the two spatial domains, the local domain is better suited for the CPOD-mrDMD stability analysis as it isolates the location of the CPOD receptivity event.
This localization is important because CPOD captures the desired forcing/stability process, but within the context of dynamics throughout the entire domain. 
In that regard, Fig.~\ref{fig:loc_glob}(a) demonstrates that the global dynamics across the jet are strongly correlated to the feedback forcing event.
Consequentially, if this global CPOD mode was used for stability analysis, the targeted instability located at the initial shear-layer would be orders of magnitude smaller than the strongly correlated impingement fluctuations at the plate, effectively making the subsequent DMD eigenvalue growth rates indeterminate. 
In other words, the local instability would be drowned out and global growth rates may not be decisively attributed to the targeted cause.
Alternatively to changing the domain size, the weighting matrix $\textbf{W}$ can be used to zero-out particular regions of the flow, as has been done previously \cite{Schmidt2019}.
The global domain is however, still well-suited for the tonal CPOD-DMD analysis, discussed next. 
Further details on how the CPOD spatial domain effects the upcoming DMD results can be found in \cite{stahl_caes2022}.





\subsection{CPOD-DMD spectral analysis}\label{secn:DMD_sub_DMD}

\begin{figure}
\centering
\includegraphics[width=1\textwidth]{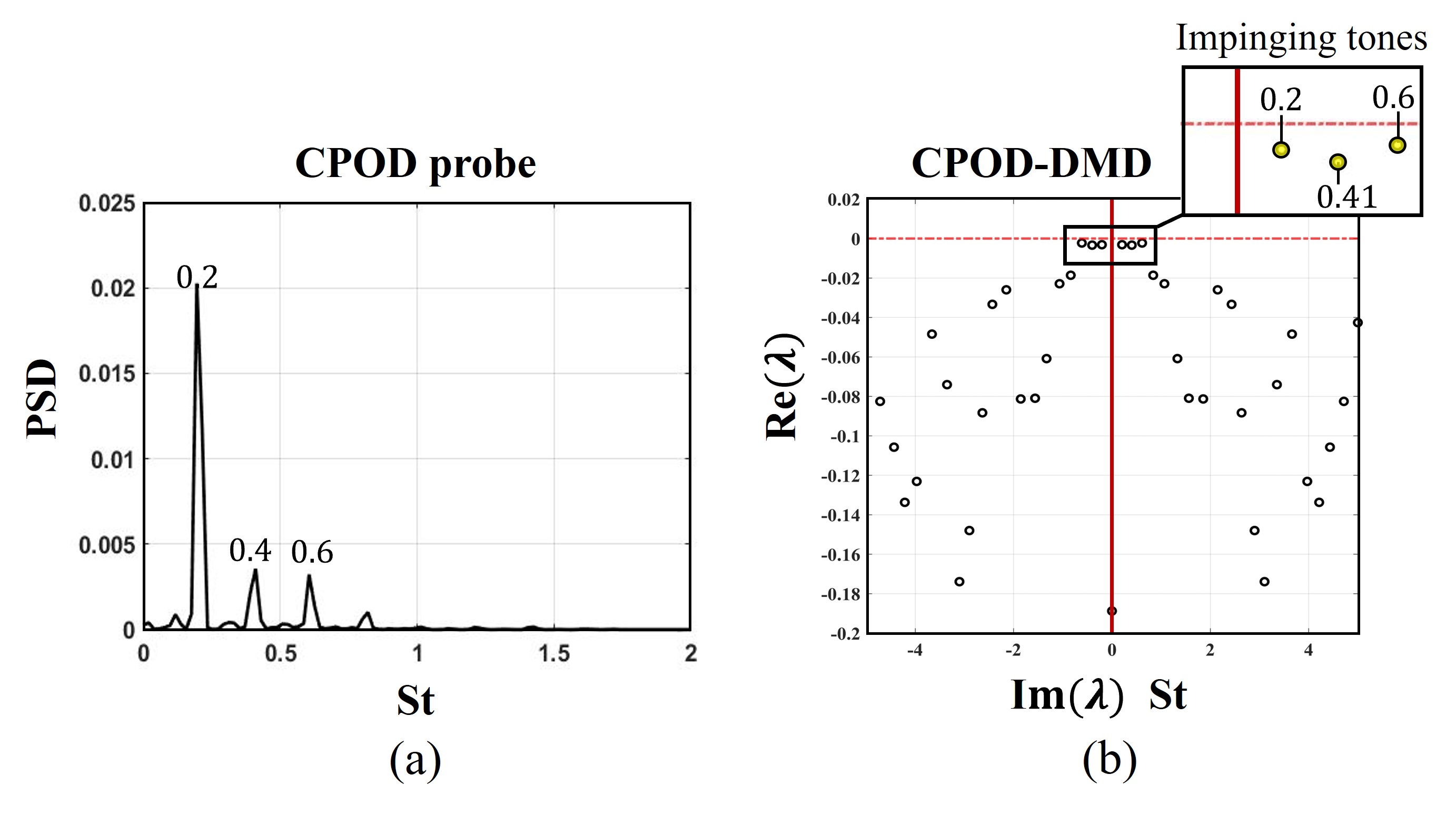}
\caption{(a) Power spectral density of the acoustic receptivity pressure probe at the nozzle showing impinging tone harmonics at $St=0.2, 0.4,$ and $0.6$. (b) CPOD-DMD eigenvalue spectra with focus on the highlighted modes at the same frequencies.}
\label{fig:SIJ_tones}
\end{figure}

DMD is now applied directly to the global CPOD mode sequence of Fig.~\ref{fig:loc_glob}(a) with the goal of elucidating spectral resonance modes.
Some CPOD-DMD modes are expected to match impinging tone harmonics, the frequencies of which are identified from the power spectral density (PSD) of the CPOD pressure signal previously shown in Fig.~\ref{fig:imping_intro}(b).
The resulting PSD is plotted in Fig.~\ref{fig:SIJ_tones}(a) and compared to the (b) CPOD-DMD eigenvalue spectrum.
Both spectra show the same impinging tone frequencies at $St=0.2, 0.4$ and $0.6$ (yellow circles in Fig.~\ref{fig:SIJ_tones}(b) cutout).
Because DMD produces as many modes as there are snapshots in the CPOD sequence, the dynamics form a relatively low rank system. 
In this example, all CPOD-DMD modes are stable with most being insignificant as their eigenvalues decay rapidly. 
The eigenvalues closest to the $Re(\lambda)=0$ stability line are oscillatory and therefore selected because they likely involve resonance behavior.

\begin{figure}
\centering
\includegraphics[width=1\textwidth]{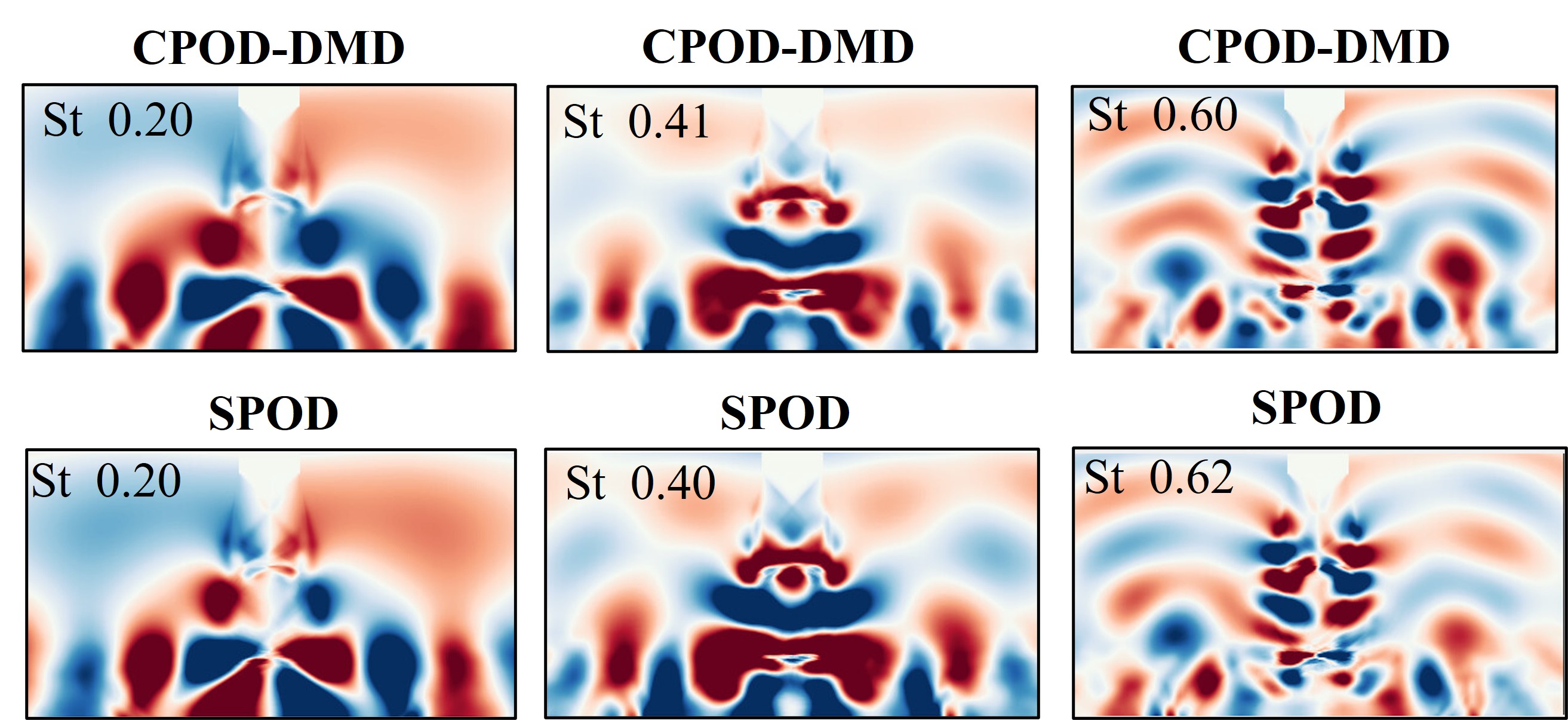}
\caption{CPOD-DMD modes of the three impinging tones compared to the leading SPOD modes calculated on equivalent window sizes ($\Delta T=40$). The CPOD-DMD extension produces nearly identical modeshapes. }
\label{fig:SPOD_compare}
\end{figure}

The three CPOD-DMD modes identified in the eigenvalue spectrum ($St=0.2, 0.4$ and $0.6$) are now presented in Fig.~\ref{fig:SPOD_compare}.
The leading SPOD modes at these frequencies are also compared, showing a nearly exact replication of the impinging tone modeshapes.
Note that for DMD to accommodate the lowest frequency tone, the CPOD time-window must be extended to at least that timescale, which for this system is $\Delta T=41$ (20 snapshots before and after the event).
In the increasing limit of $\Delta T$, CPOD event sequences overlap with each other and the CPOD-DMD framework conceptually begins to resemble the overlapping blocked structure of the SPOD algorithm \cite{Gordeyev2000,Citriniti2000}. 
Here, the SPOD modes were calculated with the same snapshot data-block size as the CPOD time window with a $50\%$ overlap.

The connection between CPOD, DMD, and SPOD is discussed more concretely. 
Revisiting the established link between DMD and SPOD \cite{Towne2018_relation}, it is well known that DMD is equivalent to a Direct Fourier Transform (DFT) if obtained from statistically-stationary, mean-subtracted snapshots \cite{Rowley2009,Chen2012}.
It is also proven that SPOD modes are optimally averaged DMD modes if obtained from an ensemble of independent flow sequences \cite{Towne2018_relation}, such as in the formulation of "exact DMD" defined in the manner of Tu \textit{et al.}~\cite{Tu2014}.
Furthermore, this parallels the general connections Tu \textit{et al.} made across DMD, Koopman spectral analysis, Linear inverse modeling (LIM) and Markov models, where DMD eigenvectors calculated from dynamics projected onto a low-dimensional subspace (derived from any of these methods for example), return the statistically most likely state at a particular frequency, which is the same optimization of variance as SPOD.
Here, the precursor CPOD consists of an ensemble of flow realizations projected onto a low-dimensional subspace, and thus satisfies the DMD-SPOD equivalency requirement, despite the CPOD mode itself not being statistically stationary.
Therefore, the combination of CPOD-DMD can approximate SPOD modes, although this relationship would likely degrade if the CPOD event time-window was shorter and more intermittently spaced (no overlap), thus capturing only partial statistics of the full system.
An interesting property was also noted regarding the use of sub-optimal CPOD modes for this impinging jet example \cite{stahl_caes2022}.
It was found that CPOD-DMD still reproduced the leading SPOD mode, as opposed to non-leading SPOD modes (which had negligible energy).

\subsection{CPOD-mrDMD cause and effect stability analysis}\label{secn:DMD_sub_mrDMD}

\begin{figure}
\centering
\includegraphics[width=1\textwidth]{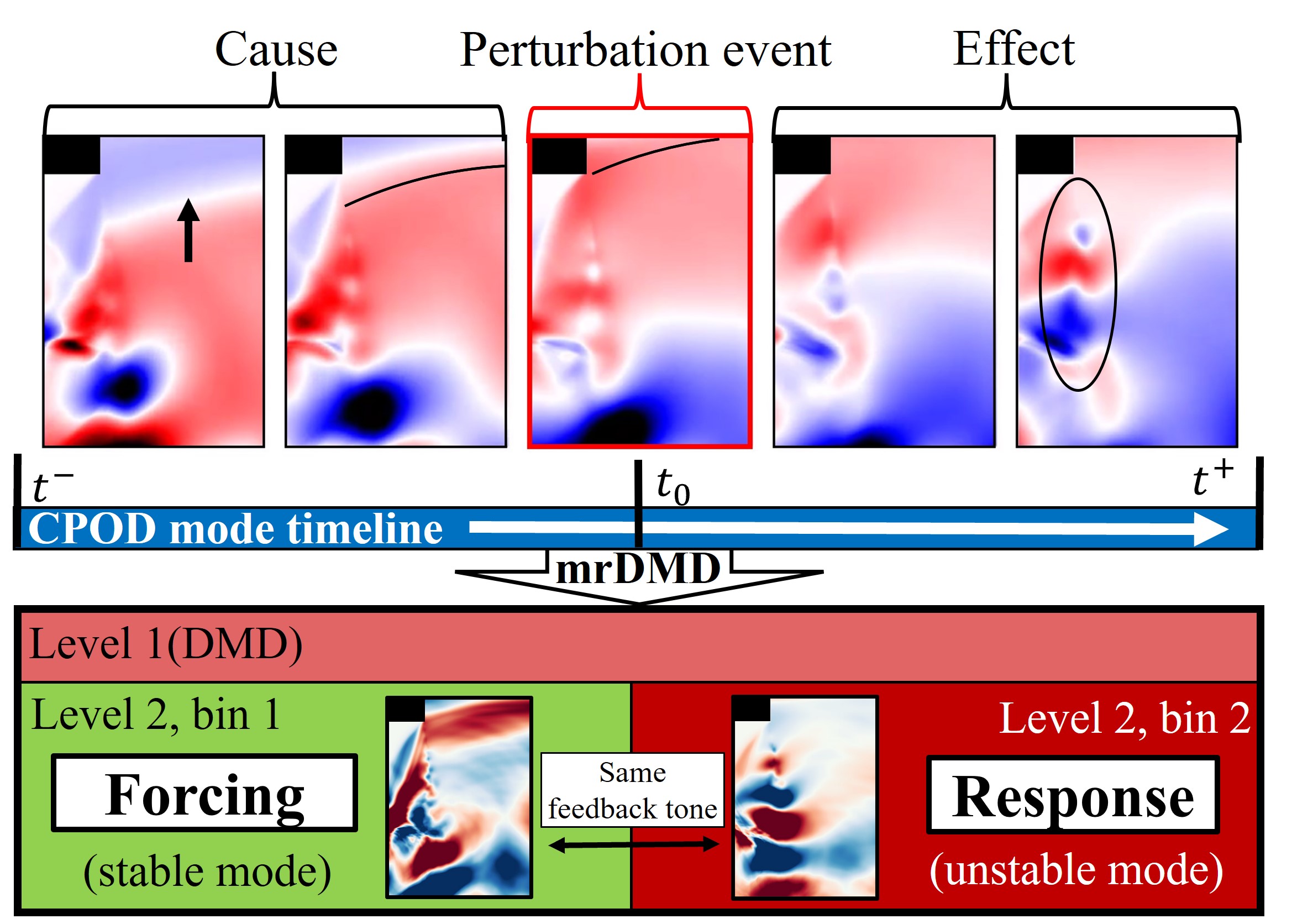}
\caption{CPOD-mrDMD framework for the cause and effect stability analysis. The CPOD mode sequence is fed into two levels of mrDMD, the second of which is partitioned into cause and effect bins, before and after the event, to identify forcing and unstable response modes.   }
\label{fig:Cause_effect}
\end{figure}

DMD is now applied to CPOD modes in a "cause and effect" framework conducive to stability analysis.
This is achieved through the multi-resolution variant, mrDMD, which provides a wavelet-like hierarchy that recursively performs the DMD algorithm to progressively smaller time-localized bins.
A diagram of the CPOD-mrDMD cause and effect stability analysis framework is illustrated in Fig.~\ref{fig:Cause_effect} using the same example of the impinging jet acoustic receptivity process.
The mrDMD time-binning is such that it segregates the CPOD timeline before and after the event.
The diagram highlights the ability of CPOD-mrDMD to isolate the causal forcing (acoustic perturbation) and effected response (shear-layer wave packet) from tonal modes.
Furthermore, the frequency and growth rate information of the CPOD-mrDMD modes gives additional insight to the dynamics of the instability.

\begin{figure}
\centering
\includegraphics[width=1\textwidth]{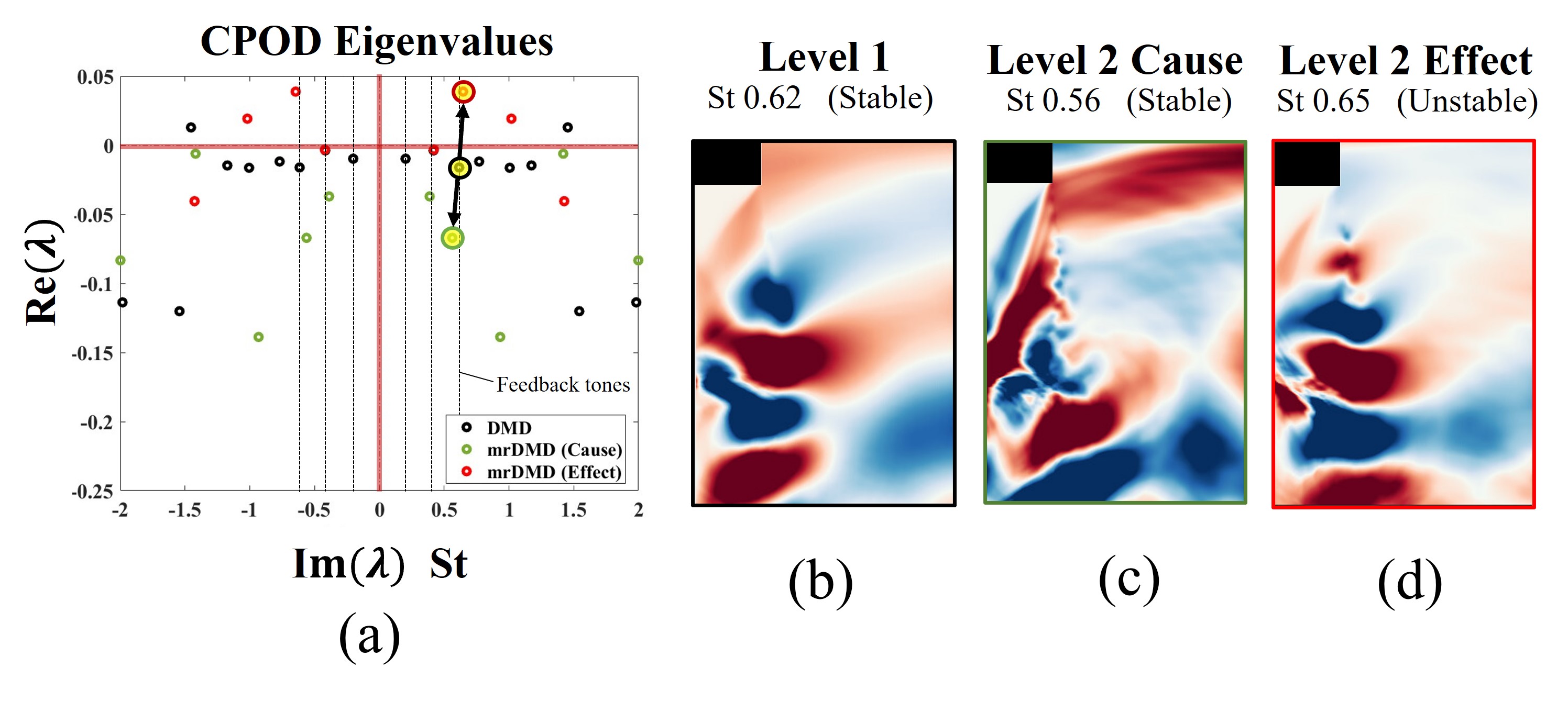}
\caption{(a) Eigenvalue spectra of all three CPOD-mrDMD mode sets. Modes corresponding to the (b) $St=0.6$ feedback tone, (c) causal acoustic forcing and (d) the resulting instability.  }
\label{fig:C_A_results}
\end{figure}

In general, the mrDMD algorithm (see Appendix~\ref{secn:appen_DMD}) can include any number of recursion levels and partitioned time-bins, removing low frequency dynamic modes with each level.
For the cause and effect analysis, a two level decomposition is sufficient, with the second level partitioned into equal "cause" and "effect" time-bins.
This produces three sets of CPOD-mrDMD modes, the eigenvalues of which can be compared to understand how the event induces changes to the flow.
Returning to the impinging jet example, Fig.~\ref{fig:C_A_results}(a) shows the CPOD-mrDMD eigenvalues for each level and bin.
In level~1, which spans the entire CPOD sequence, the three dominant impinging tones $St=0.2, 0.4$ and $0.6$ are still present.
At level~2, the "cause" eigenvalue spectra is completely stable while the "effect" spectra undergoes a shift upward including multiple unstable modes. 
All modes corresponding to the highlighted $St=0.6$ impinging tone are plotted in Fig.~\ref{fig:C_A_results} (b-d), demonstrating how the agglomerated feedback cycle of level~1 (b) can be separated into its constituent forcing (c) and unstable response (d) components, including the shift in time-local frequencies.
Furthermore, the inconspicuously small $St=0.6$ impinging tone harmonic in the PSD of Fig.~\ref{fig:SIJ_tones}(a) is discovered to play a large role in perpetuating the resonance instability.

Finally, some comments are made on choice of CPOD parameters (temporal window size, spatial domain, and CPOD mode selection) that should be tailored to a localized analysis for best results.
Regarding the temporal window size $\Delta T$, the window should be minimized to just capture the phenomenon of interest, as relative growth rates tend to diminish with a larger window. 
This philosophy is the opposite of the previous CPOD-DMD spectral analysis that had the goal of mimicking SPOD with larger windows. 
A smaller sized or weighted spatial domain is also preferred for stability analysis for the reasons discussed in \ref{secn:DMD_sub_SIJ}.
Note, that an alternative CPOD-DMD stability analysis, without multiresolution cause and effect localization, could achieve similar results with the commensurate time-window following the event. 
While only the first CPOD mode was used in the impinging jet example, non-leading CPOD modes could test how lower energy dynamics are involved in the instability process.
A strategy where multiple CPOD modes are weighted by singular values and summed together, could be effective as DMD can filter the additional contributions \cite{stahl_caes2022}. 

\section{CPOD for event prediction}\label{secn:predict}

\begin{figure}
\centering
\includegraphics[width=1\textwidth]{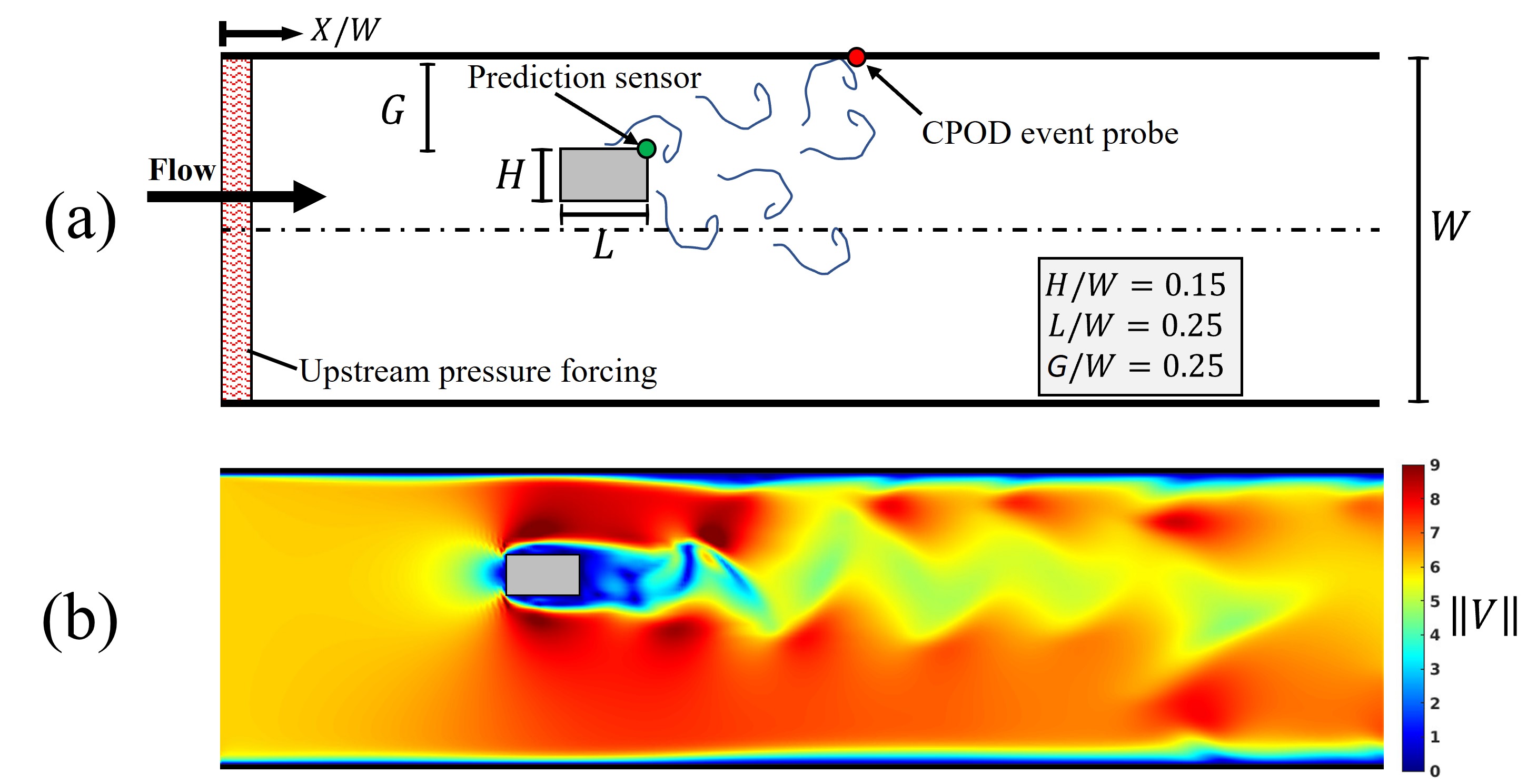}
\caption{(a) Configuration of the bluff body offset in a channel and (b) the instantaneous velocity magnitude. The CPOD event probe characterizes the impingement of wake structures and the prediction sensor actively measures the flow and compares with the CPOD mode to predict future events. }
\label{fig:predict_demo}
\end{figure}

A strategy is presented that correlates CPOD modes with an active sensor to predict high fluctuation events in the near-term future.
The system under consideration is a bluff body offset in a narrow channel, such that wake structures impinge on the top surface of the channel, as represented in Fig.~\ref{fig:predict_demo}(a).
For a simple event prediction strategy, two points in the flow are specified, the active sensor placed on the bluff body, and the CPOD correlation probe located on the upper wall which is used to characterize the wake impingement events. 
The fundamental idea is that the recent time-history of the upstream sensor is constantly being compared to the CPOD mode at the same location; when the measured flow pattern aligns with the CPOD trajectory, an event on the wall is likely to happen at a predefined future time horizon. 
The CPOD prediction framework can be partitioned into three steps within a simulation or experiment: 1) the initial data-gathering of repeated events that lasts until sufficient statistics are collected, 2) the in-situ CPOD analysis, and 3) the active flow sensing and comparison with the CPOD mode to predict future events that lasts for the remainder of the time.

The configuration details of the undeveloped channel flow and a description of the wake impingement events are first discussed.
The simulation models the 2D incompressible Navier-Stokes equations with a Reynolds number of $6000$, based on the channel width $W=1m$ and inlet velocity $U_{\infty}=6m/s$.
The blockage effect of the rectangular body ($H/W=0.15$, $L/W=0.25$), offset from the channel center-line, induces faster flow above the body relative to below, as visualized in Fig.~\ref{fig:predict_demo}(b) with the instantaneous velocity magnitude. 
This asymmetry, combined with the boundary layer interactions on the channel wall, distort the typical periodic wake shedding behavior.
Additional non-linearity is introduced to the system by a spatially and temporally stochastic pressure forcing (uniformly random distribution with a max fluctuation of $5\%$ dynamic inlet pressure) applied upstream of the body.
All together, this results in irregular sized vortices that intermittently impinge on the upper wall.

\begin{figure}
\centering
\includegraphics[width=1\textwidth]{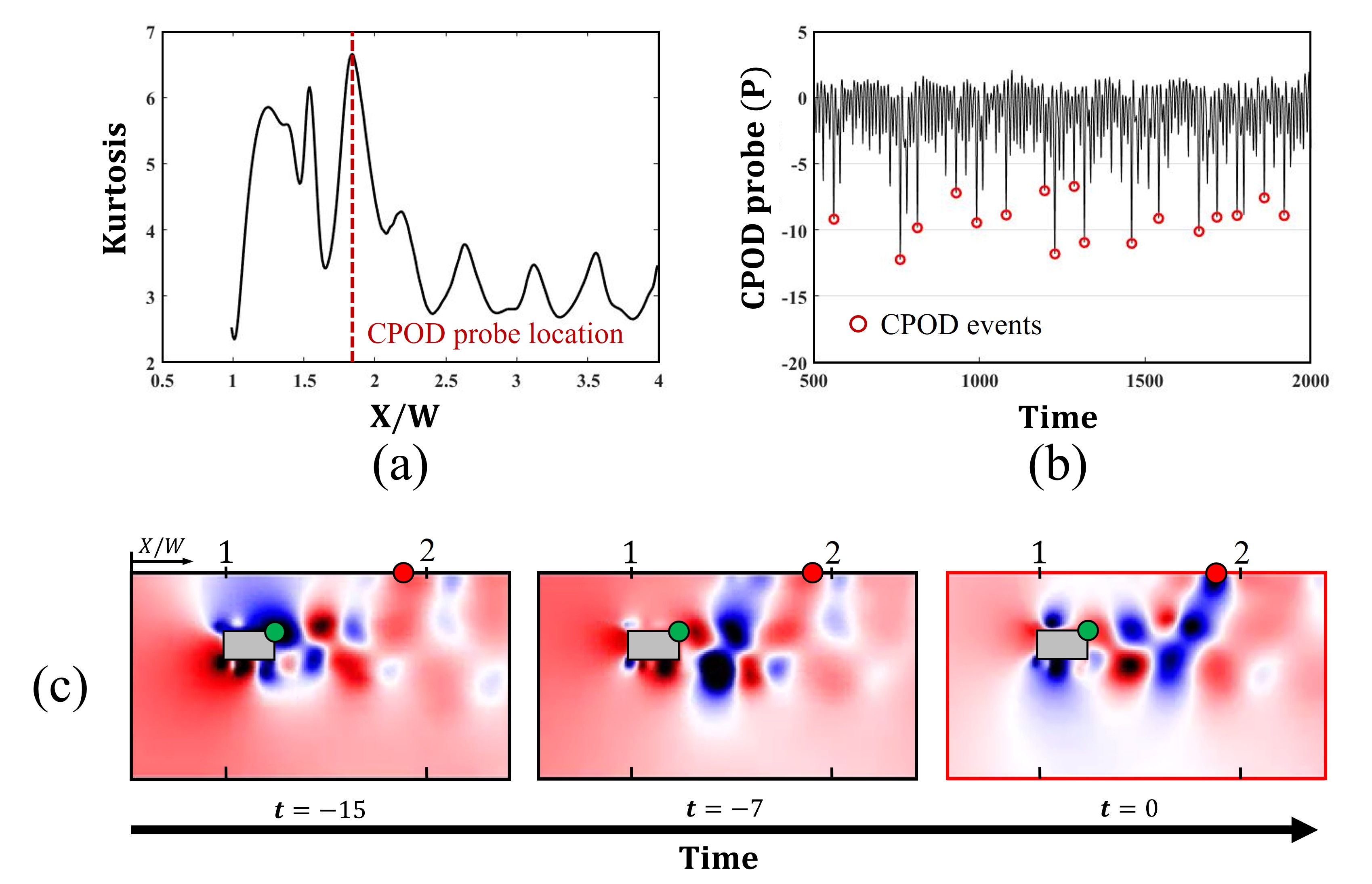}
\caption{ (a) Kurtis of pressure across the upper channel wall is used to identify the optimal CPOD probe location. (b) The pressure signal at the peak kurtosis (red dot, $X/W=1.88$) identifies the extreme wake impingement events. (c) Snapshots from the first CPOD mode corresponding to these events, progressing in time from left to right.}
\label{fig:predict_signal}
\end{figure}

After phase 1) of the simulation has collected enough of these event realizations, the CPOD mode is calculated.
First, the question of choosing the optimal CPOD probe location arises to best capture the event.
In order to find this point, the kurtosis of the pressure along the upper surface is calculated and plotted in Fig.~\ref{fig:predict_signal}(a).
Kurtis represents the "tailedness" or level of extreme fluctuations in a signals probability distribution, which peaks at $X/W=1.88$ on the wall (red line).
A sample of the signal at this location is plotted in Fig.~\ref{fig:predict_signal}(b) to exhibit the extreme negative pressure fluctuations, which are therefore identified as CPOD events (red circles).
In phase 1) of the simulation, $25$ events were identified, ensuring that that no two events are captured in the same realization window.
In phase 2), the corresponding first CPOD mode captures $20\%$ percent of the energy, and is shown in Fig.~\ref{fig:predict_signal}(c) progressing in time from left to right.
A portion of the CPOD mode sequence leading up to the event ($t=0$) reveals the asymmetric, blue coherent structures traveling from the trailing edge of the body to the channel wall.
Logically, the live sensor that measures the flow and is compared to the CPOD mode, is placed at this trailing edge location (green circle).

\begin{figure}
\centering
\includegraphics[width=.95\textwidth]{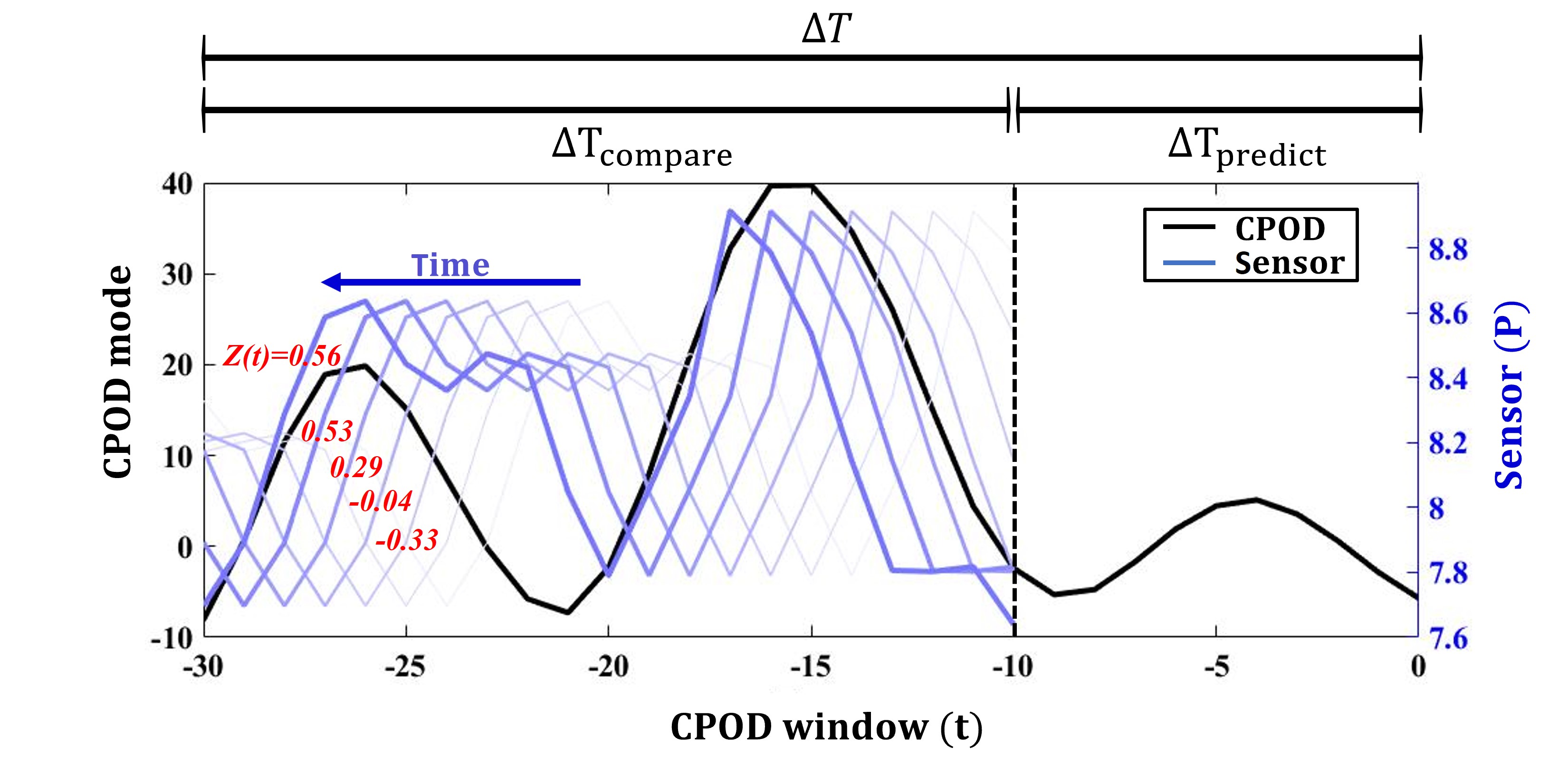}
\caption{ Sensor comparison to the CPOD mode at the same location. The continually updating sensor measurement history is correlated ($Z(t)$) over the period $\Delta T_{compare}$ to predict events $\Delta T_{predict}$ in the future. }
\label{fig:sig_correlate}
\end{figure}

The active sensor to CPOD mode comparison requires considerations of CPOD window size and the prediction time horizon.
Regarding the CPOD window size $\Delta T$, the time before the event $t^{-}$ necessarily must be atleast long enough to capture the entire path of the coherent structure traveling from the sensor location to the CPOD correlation point, and ideally longer to recognize preceding flow patterns at the sensor location.
The CPOD window limit after the event, $t^{+}$, is less important and thus minimized. 
These parameters can be adjusted heuristically in-situ as the CPOD calculation is very fast and easily iterated upon. 
The next most important factor to consider is the prediction time horizon, $\Delta T_{predict}$, which also determines the length of sensor time-history that is compared to the CPOD mode, $\Delta T_{compare}$, by the relation
\begin{equation}
    \Delta T_{compare}=\Delta  T - \Delta T_{predict}.
\end{equation}
In this example, these times are $\Delta T=30$, $\Delta T_{predict}=10$, and $\Delta T_{compare}=20$.
Their connection to each other is further illustrated in  Fig.~\ref{fig:sig_correlate} with the sensor to CPOD mode signal comparison.
Here, the continuously updating pressure sensor time-history (right Y-axis) is shifted back by $\Delta T_{predict}$ and correlated to the constant CPOD mode values (Left Y-axis) at the same spatial location.
At every time-step in phase 3), the overlapping portions of these two signals are correlated, producing the coefficient $Z(t)$.
The utility of this coefficient being that when the measured flow at the sensor is well-correlated to the CPOD mode, the wake impingement event is likely to occur at the CPOD event location at the wall, $\Delta T_{predict}$ time-steps into the future.

To test this, a prediction likelihood estimate is quantified based on the sensor correlation $Z(t)$ from $[-1, 1]$. 
In this example, the prediction certainty is simply converted to a percentage $\widetilde{Z}(t)$, defined by 
\begin{equation}
    \widetilde{Z}=\frac{Z+1}{2}\times100
\end{equation}.

In general, more complicated prediction certainty functions can be developed to improve accuracy.
Possibilities might include time-lag cross-correlations, phase synchronization metrics or be a function of multiple weighted sensors and CPOD modes.
The prediction accuracy also has a clear dependence on the three timescale parameters which can be optimized by iterating over phases 2) and 3).
As with all turbulent fluid systems, correlations and predictions are expected to degrade over longer time horizons.

\begin{figure}
\centering
\includegraphics[width=1\textwidth]{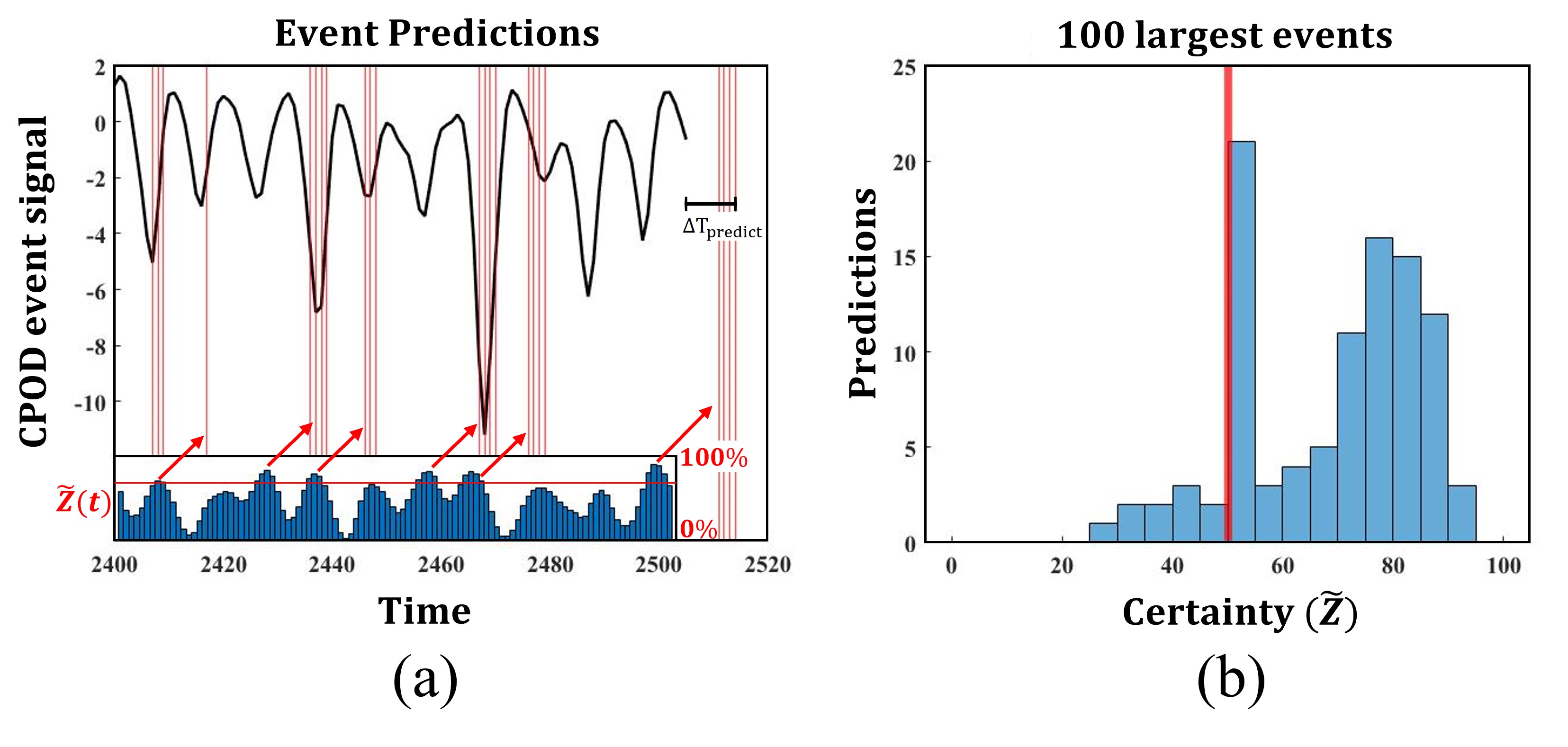}
\caption{(a) Sample pressure signal at the event location and the prediction certainty percentage. Red lines are plotted $\Delta T_{predict}$ time-steps into the future when $\widetilde{Z}(t) > 70\%$. (b) Histogram of of the prediction certainty preceding the $100$ largest events.   }
\label{fig:predict_report}
\end{figure}

The CPOD event prediction results are now summarized in Fig.~\ref{fig:predict_report}(a) with a sample of the pressure signal at the channel wall to demonstrates how the certainty percentage $\widetilde{Z}(t)$ predicts future events, and (b) a post-facto test of the prediction certainty distribution that preceded the $100$ largest events.
Examining Fig.~\ref{fig:predict_report}(a), vertical red lines are marked when $\widetilde{Z}(t) > 70\%$ at $\Delta T_{predict}$ time-steps into the future. 
These predicted instances generally come to fruition and coincide with the extreme pressure events, but also occasionally predict minor fluctuations.
More quantitatively, the prediction estimations preceding the largest $100$ fluctuation events (gathered at $\Delta T_{predict}$ time steps earlier) are plotted in the histogram of Fig.~\ref{fig:predict_report}(b).
The distribution shows a clear bias that correctly predicts when an event is more likely to happen with the median estimate being $72\%$.
However, a significant outlier of $20$ events have an inconclusive $50-55\%$ certainty, possibly due to dynamics in non-leading CPOD modes that are not correlated with the sensor. 
While certainty cannot be guaranteed for any nonlinear chaotic system, the causal insights extracted from CPOD can tilt the favor of a successful prediction strategy and potentially give an edge to future control strategies.


\section{Conclusion}

This paper expounds upon Conditional space-time Proper Orthogonal Decomposition (CPOD) by exploring connections to other common data-driven decomposition methods and proposing novel CPOD extensions and modeling applications.
First, the theoretical foundation of CPOD is outlined as formulated by Schmidt \textit{et al.} \cite{Schmidt2019}.
Essentially, CPOD is an ensemble of conditionally selected event realizations that are decomposed to optimally maximize the variance across a set modes; these modes are a function of space and time, and orthogonal in the space-time inner product.
The "events" are in general arbitrary, but in this work include intermittent turbulent spots in a transitioning supersonic boundary layer, supersonic inlet engine unstart "buzz", aeroacoustic feedback mechanisms in a supersonic impinging jet, and bluff-body wake-flow interactions with a channel wall.
Apart from these complex fluid systems, the CPOD methodology is instantiated with the chaotic Lorenz system, with discussion on the closely related ensemble average.  
Practical considerations when performing CPOD, including the event sequence temporal window size, spatial domain, and event selection criteria are also discussed.

A typical CPOD fluid analysis is demonstrated on the example of a Mach $2.5$ transitioning supersonic boundary layer. 
The CPOD modes are able to deduce the transient mechanisms that are associated with the turbulent spots, including preceding streaks and resulting acoustic radiation.
The effect of the conditional criteria is also tested and compared to space-only POD, demonstrating how oversampling can lower the rank of the modeled dynamics based on arguments of time-delay embedding.
A similar CPOD analysis is presented for an unstarted supersonic inlet and compression spike buzz.
The goal being to demonstrate how CPOD can easily be applied as a video post-processing tool for experimental data where events are selected from pixel intensity. 
Results show the shock-wave propagation and boundary-layer separation dynamics are easily separated and visualized across leading CPOD modes.

Next, CPOD is extended with dynamic mode decomposition (DMD), applied directly to the transient CPOD mode sequence.
The ability of CPOD-DMD to filter the conditionally selected dynamics by frequency and growth rate yields a versatile tool for spectral and stability applications. 
Two strategies are proposed: 1) a spectral CPOD-DMD analysis that can exactly reproduce SPOD modes, and 2) a multiresolution framework, CPOD-mrDMD, that yields a "cause and effect" stability analysis. 
Both DMD extensions are applied to a supersonic impinging jet to study 1) the resonance tones generated from the aeroacoustic feedback mechanism, and 2) the local receptivity of the causal acoustic feedback forcing and resulting shear-layer instability.
Regarding the former method, it is shown that a longer CPOD time-window nullifies the DMD eigenvalue growth rates and the CPOD-DMD methodology begins to resemble the overlapping ensembled block structure of the SPOD algorithm \cite{Towne2018_relation}.
The later approach, CPOD-mrDMD, offers a time-local analysis that can separate the causal forcing and resulting effects on flow stability, as growth rates become relevant over shorter time-horizons.
This framework is advantageous as a diagnostic tool to examine large fluctuation events.
This is because the "natural" flow conditions that produce the forcing event and nonlinear response are inherently captured in the CPOD mode; whereas other stability methods artificially prescribe forcing perturbations around a laminar or mean base-state and calculate a linearized response.

Finally, the potential for CPOD as a tool for prediction and control applications is appreciated through an example of wake-flow behind a bluff body in a channel.
Here, the configuration is offset in the channel with random forcing upstream of the body to induce non-linearity, such that wake-flow structures intermittently impinge on the channel wall.
A relatively simple prediction scheme is procured where a CPOD mode that characterizes the impingement event is constantly being correlated to the immediate time-history of an active sensor on the bluff body.
When the measured flow aligns with the CPOD mode trajectory, an impingement event at the wall is likely to occur at an estimated time in the future. 
The successful results suggest that CPOD modes are a viable model for prediction strategies, with much room for future improvement coming from the developments of the flow control community.


\appendix\section{Appendix: Dynamic Mode Decomposition} \label{secn:appen_DMD}

The methodology and algorithm behind DMD is now summarized.
Further discussion on the theory and implementation of DMD can be found in references \cite{Rowley2009,Schmid2010,Tu2014,Chen2012}.
Broadly, the goal of DMD is to take a sequence of snapshots and produce a linear operator that best-fits the non-linear evolution of flow from one snapshot to the next. 
From this operator, an eigenvalue decomposition provides a set of dynamic modes with associated frequencies and growth rates. 
The connection to CPOD is made by providing output CPOD modes as an input for DMD.
For this, the CPOD mode $\Phi(\textbf{x},r)$ is rearranged in an $N \times M$ matrix, where $N$ is the number of DOFs in each snapshot and $M$ is the number of snapshots in the CPOD mode across $\Delta T$,

\begin{equation}
    \boldsymbol{\Phi}= [{\phi}(\textbf{x},t^{-}) \hspace{0.05in}... \hspace{0.05in}{\phi}(\textbf{x},t_{o}) \hspace{0.05in}... \hspace{0.05in}{\phi}(\textbf{x},t^{+})],
\end{equation}
 ($M=t^{-}+t^{+}+1$).
The approximated linear operator, \textbf{A}, that best fits the dynamics from one snapshot to the next is
\begin{equation}
    {\phi}_{i+1}=\textbf{A}{\phi}_{i}.
\end{equation}
This $\textbf{A}$ matrix is a very large $N \times N$ matrix which can be intractable to work with.
The DMD algorithm instead uses a low-rank approximation $\tilde{\textbf{A}}$ projected from a POD subspace.
This is accomplished by considering the snapshot matrix $\boldsymbol{\Phi}$ separated into two shifted snapshot matrices $\boldsymbol{\Phi}_{1}$ and $\boldsymbol{\Phi}_{2}$ as follows 
\begin{equation}
    \boldsymbol{\Phi}_{1}= [\phi(\textbf{x},t^{-})...{\phi}(\textbf{x},t_o)],
\end{equation}

\begin{equation}
    \boldsymbol{\Phi}_{2}= [{\phi}(\textbf{x},t_o+1)...{\phi}(\textbf{x},t^{+})].
\end{equation}
The data matrix $\boldsymbol{\Phi}_{1}$ is then orthogonalized using the singular value decomposition
\begin{equation}
    \boldsymbol{\Phi}_{1}=\mathbf{U\Sigma V^{*}}
\end{equation}
where * denotes the conjugate transpose. 
The full sized matrix $\textbf{A}$ would be computed as:
\begin{equation}
      \mathbf{A}=\boldsymbol{\Phi}_{2}  \mathbf{V\Sigma^{-1}U^{*}}
\end{equation}
However, by substituting the low-rank projected modes $U$, the reduced matrix $\tilde{\textbf{A}}$ can be calculated as
\begin{equation}
      \mathbf{\tilde{A}}=  \mathbf{U^{*}AU}=  \mathbf{U^{*}}\boldsymbol{\Phi}_{2}  \mathbf{V\Sigma^{-1}}
\end{equation}
The reduced $M x M$ linear operator $\mathbf{\tilde{A}}$ is then eigendecomposed as
\begin{equation}
    \mathbf{\tilde{A}W}=\mathbf{W\Lambda},
\end{equation}
where the columns of $\mathbf{W}$ are eigenvectors and $\mathbf{\Lambda}$ is a diagonal matrix containing the corresponding eigenvalues $\lambda$.
The DMD modes $\boldsymbol{\Psi}$ are then projected back onto the full size domain:

\begin{equation}
    \boldsymbol{\Psi}=\boldsymbol{\Phi_{2}}\mathbf{V\Sigma^{-1}W}
\end{equation}

The eigenvalue spectrum corresponding to each DMD mode is complex and can be rewritten as $\omega_k=\log(\lambda_k)/\Delta t$, where $\Delta t$ is the time between snapshots and $k$ is the DMD mode index. 
The real and imaginary parts of $\omega$ represent the growth rate and frequency of the DMD mode respectively. 
Unstable DMD modes are characterized by a positive growth rate and stable modes by a negative groth rate.
This is apparent by considering the DMD reconstruction of the input time dynamics
\begin{equation}
   \boldsymbol{\Phi}(t) \approx \sum_{k=1}^{M}\psi_k \exp(\omega_k t)b_k = \boldsymbol{\Psi}\exp(\boldsymbol{\Omega}t)\textbf{b}
\end{equation}
where $b_k$ is the column vector containing the the initial amplitude of the mode, calculated by the projection of modes onto the first input snapshot
\begin{equation}
   \textbf{b}=\Psi^{\dagger}\phi(t^{-}).
\end{equation}

The multi-resolution extension of DMD (mrDMD) \cite{Kutz2015} recursively performs this algorithm in a hierarchical wavelet transform framework.
The compartmentalization of the input snapshot matrix into time bins $\mathscr{j}$ makes resulting mrDMD modes ideal for elucidating temporally local dynamics and separation of timescales, as low-frequency modes are removed with each recursive level of decomposition $\mathscr{l}$.
The CPOD-mrDMD cause and effect stability framework is created by using a $\mathscr{l}=2$ level decomposition.
Level~$1$ is simply the DMD algorithm applied over the entire CPOD sequence, while level~$2$ is partitioned into $\mathscr{j}=2$ bins.
Bin $1$, dubbed "Cause", encompasses snapshots $t^-$ to $t_o$ and bin $2$, "Effect", is from $t_o$ to $t^+$.
In total, there are three sets of mrDMD modes $\Psi^{\mathscr{l},\mathscr{j}}_{k}$ with corresponding eigenvalue spectra for each ${\tilde{\textbf{A}}}^{\mathscr{l},\mathscr{j}}$ calculated.

All modes with eigenvalues ($|\omega_k| \leq \rho$) in the level~1 bin are considered slow modes.
These modes are reconstructed and subtracted from the snapshot matrix to remove the low frequency time dynamics before moving onto level 2.
In the CPOD-mrDMD analysis, $\rho$ follows the convention of Ref.~\cite{mrDMD.ch5} and is chosen to select slow modes with fewer than two cycles of oscillation within the sampling window. 

The reconstruction of the cause and effect CPOD dynamics in the mrDMD framework is given as
\begin{equation}
   \boldsymbol{\Phi}(t) \approx \sum_{\mathscr{l}=1}^{2} \sum_{\mathscr{j}=1}^{\mathscr{j}_{\text{max}}}   \sum_{k=1}^{m} {f}^{\mathscr{l},\mathscr{j}}(t)b_{k}^{\mathscr{l},\mathscr{j}}\phi_{k}^{\mathscr{l},\mathscr{j}}\exp{(\omega_{k}^{\mathscr{l},\mathscr{j}}t)},
\end{equation}
where $f$ is an indicator function equal to either zero or one depending on if the modes $\phi_{k}^{\mathscr{l},\mathscr{j}}$ lie within the current time bin.
Selecting DMD modes with a common frequency from each time bin provides insights into the local stability dynamics before and after the event.

\section*{Acknowledgements}
This work was performed in part under the sponsorship of the Office of Naval Research (Contract N00014-18-1-2506) with Dr. D. Gonzalez serving as Project Monitor. The views and conclusions contained herein are those of the authors and do not represent the opinion of the Office of Naval Research or the U.S. government. Computational resources were provided by the DoD High Performance Computing Modernization Program as well as the Ohio Supercomputer Center. Several figures were made using FieldView software with licenses obtained from the Intelligent Light University Partnership Program. 


\bibstyle{unsrt}
\bibliography{CPOD_manuscript.bib}



\end{document}